\documentclass[aps,prl, twocolumn,superscriptaddress, nobibnotes]{revtex4-1}
\usepackage{graphicx}
\usepackage[hidelinks]{hyperref} 
\usepackage{bm}
\usepackage{amsmath}
\usepackage{bbold}

\graphicspath{{./}}

\begin{document}


\title{Line-Graph Approach to Spiral Spin Liquids} 
\thanks{This manuscript has been authored by UT-Battelle, LLC under Contract No. DE-AC05-00OR22725 with the U.S. Department of Energy.  The United States Government retains and the publisher, by accepting the article for publication, acknowledges that the United States Government retains a non-exclusive, paid-up, irrevocable, world-wide license to publish or reproduce the published form of this manuscript, or allow others to do so, for United States Government purposes.  The Department of Energy will provide public access to these results of federally sponsored research in accordance with the DOE Public Access Plan (http://energy.gov/downloads/doe-public-access-plan).}

\renewcommand*{\thefootnote}{\arabic{footnote}}

\author{Shang Gao}
\email[]{sgao.physics@gmail.com}
\affiliation{Neutron Scattering Division, Oak Ridge National Laboratory, Oak Ridge, Tennessee 37831, USA}
\affiliation{Materials Science \& Technology Division, Oak Ridge National Laboratory, Oak Ridge, Tennessee 37831, USA}

\author{Ganesh Pokharel}
\affiliation{Materials Science \& Technology Division, Oak Ridge National Laboratory, Oak Ridge, Tennessee 37831, USA}
\affiliation{Department of Physics \& Astronomy, University of Tennessee, Knoxville, Tennessee 37996, USA}

\author{Andrew F. May}
\affiliation{Materials Science \& Technology Division, Oak Ridge National Laboratory, Oak Ridge, Tennessee 37831, USA}

\author{Joseph A. M. Paddison}
\affiliation{Materials Science \& Technology Division, Oak Ridge National Laboratory, Oak Ridge, Tennessee 37831, USA}

\author{Chris Pasco}
\affiliation{Materials Science \& Technology Division, Oak Ridge National Laboratory, Oak Ridge, Tennessee 37831, USA}

\author{Yaohua Liu}
\affiliation{Neutron Scattering Division, Oak Ridge National Laboratory, Oak Ridge, Tennessee 37831, USA}

\author{Keith M. Taddei}
\affiliation{Neutron Scattering Division, Oak Ridge National Laboratory, Oak Ridge, Tennessee 37831, USA}

\author{Stuart Calder}
\affiliation{Neutron Scattering Division, Oak Ridge National Laboratory, Oak Ridge, Tennessee 37831, USA}

\author{David G. Mandrus}
\affiliation{Materials Science \& Technology Division, Oak Ridge National Laboratory, Oak Ridge, Tennessee 37831, USA}
\affiliation{Department of Physics \& Astronomy, University of Tennessee, Knoxville, Tennessee 37996, USA}
\affiliation{Department of Material Science \& Engineering, University of Tennessee, Knoxville, Tennessee 37996, USA}

\author{Matthew B. Stone}
\affiliation{Neutron Scattering Division, Oak Ridge National Laboratory, Oak Ridge, Tennessee 37831, USA}

\author{Andrew D. Christianson}
\affiliation{Materials Science \& Technology Division, Oak Ridge National Laboratory, Oak Ridge, Tennessee 37831, USA}

\date{\today}


\begin{abstract}
Competition among exchange interactions is able to induce novel spin correlations on a bipartite lattice without geometrical frustration. A prototype example is the spiral spin liquid, which is a correlated paramagnetic state characterized by sub-dimensional degenerate propagation vectors. Here, using spectral graph theory, we show that spiral spin liquids on a bipartite lattice can be approximated by a further-neighbor model on the corresponding line-graph lattice that is non-bipartite, thus broadening the space of candidate materials that may support the spiral spin liquid phases. As illustrations, we examine neutron scattering experiments performed on two spinel compounds, ZnCr$_2$Se$_4$ and CuInCr$_4$Se$_8$, to demonstrate the feasibility of this new approach and expose its possible limitations in experimental realizations.

\end{abstract}

\maketitle

\textit{Introduction.}--- A spiral spin liquid (SSL) is an exotic correlated paramagnetic state of \textit{sub-dimensional} degeneracy, meaning that the propagation vectors $\bm{q}$ of the ground states form a continuous manifold, or spiral surface, in a dimension that is reduced from the original system~\cite{bergman_order_2007, lee_theory_2008, mulder_spiral_2010,zhang_exotic_2013, niggemann_classical_2019, attig_classical_2017, balla_degenerate_2020, balla_affine_2019, lee_theory_2008, chen_quantum_2017, yao_generic_2021, huang_versatility_2021, buessen_quantum_2018, liu_featureless_2020, iqbal_stability_2018}. 
Similar to geometrically frustrated magnets~\cite{bramwell_spin_2001, balents_spin_2010}, a SSL may host topological spin textures~\cite{gao_spiral_2017,gao_fractional_2020,shimokawa_multiple_2019} and quantum spin liquid states~\cite{mulder_spiral_2010, zhang_exotic_2013, niggemann_classical_2019, balz_physical_2016, biswas_semi_2018, pohle_theory_2021}. What differentiates a SSL from a conventional frustrated magnet is the sub-dimensional degeneracy, which induces highly distinctive dynamics since the spins are confined to fluctuate collectively as nonlocal spirals~\cite{yan_low_2021}. Recent calculations on a square lattice reveal that the low-energy fluctuations in a SSL may behave as topological vortices in momentum space~\cite{yan_low_2021}, leading to an effective tensor gauge theory with unconventional fracton quadrupole excitations that are deeply connected to theories of quantum information, elasticity, and gravity~\cite{pretko_sub_2017, pretko_emergent_2017, pretko_fracton_2018, yan_hyperbolic_2019, nandkishore_fractons_2019, pretko_fracton_2020}.

To date, bipartite lattices have been the primary avenue through which SSLs are studied. This is because the ground state degeneracy on a bipartite lattice can be exact, so that all spin spirals with $\bm{q}$ over the spiral surface have exactly the same energy~\cite{niggemann_classical_2019}. Although this degeneracy stabilizes the SSL down to very low temperatures~\cite{bergman_order_2007, mulder_spiral_2010}, it also imposes a strong constraint on real materials because most of the known bipartite-lattice compounds are dominated by the nearest-neighbor interactions $J_1$~\cite{tristan_geometric_2005, suzuki_melting_2007, krimmel_spin_2009, matsuda_disordered_2010, macdougall_kinetically_2011, zaharko_spin_2011, nair_approaching_2014, macdougall_revisiting_2016, ge_spin_2017, chamorro_frustrated_2018, tsurkan_on_2021, haraguchi_frustrated_2019, abdeldaim_realizing_2020, otsuka_canting_2020, wessler_observation_2020}. Even for the established model compounds where the second-neighbor interactions $J_2$ are relatively strong~\cite{gao_spiral_2017, guratinder_magnetic_2022, gao_spiral_2021}, the degeneracy over the spiral surface is only approximate due to the existence of further perturbations~\cite{iqbal_stability_2018}. This degeneracy lifting results in an approximate SSL state at elevated temperatures where thermal fluctuations overcome the slight energy difference among the spirals. 

Inspired by recent density functional theory (DFT) calculations for the breathing pyrochlore lattice compounds~\cite{ghosh_breathing_2019}, here we seek the realization of an approximate SSL, \textit{i.e.} a SSL with an approximate degeneracy, beyond the bipartite lattices. According to the Luttinger-Tisza theory~\cite{luttinger_theory_1946,luttinger_note_1951}, the degeneracy of a SSL model is encoded in the minimum manifold of the interaction matrix. Using graph theory~\cite{knauer_algebraic_2019, kollar_line_2020}, we show that the $J_1$-$J_2$ model on a bipartite lattice shares the same minimum manifold with a $J_1$-$J_3$ model on the corresponding line-graph lattice, where $J_3$ denotes the third-neighbor interaction. Thus an approximate SSL state is achieved in the latter case when $J_3$ is sufficiently strong, which greatly expands the range of materials that may support a SSL state. This line-graph approach to SSL is vetted through neutron scattering experiments performed on two Cr-based chalcogenide spinels ZnCr$_2$Se$_4$ and CuInCr$_4$Se$_8$.

\textit{Line-graph approach.}--- Our starting point is a Heisenberg model on a $l$-regular lattice, where $l$ counts the number of the nearest-neighbor (NN) sites. In the presence of a uniform NN exchange interaction $J_1$, the coupled spins form a undirected graph $G = (V,D)$, with $V$ denoting the set of vertices (\textit{i.e.} the spin sites) and $D$ denoting the set of edges (\textit{i.e.} the NN bonds). Two vertices $i$ and $j$ are called \textit{adjacent} if the graph contains an edge $e = \{i,j\}$, and the adjacency matrix is defined as $A(G)_{ij} = 1$ for $\{i,j\}\in D$ and 0 otherwise. Following this definition, the spin Hamiltonian, $\mathcal{H} = J_1\sum_{\langle ij\rangle}\bm{S}_i\cdot \bm{S}_j$ with $\langle ij\rangle$ denoting the NN bonds, can be expressed through the adjacency matrix as $\mathcal{H}=\frac{1}{2}J_1 A(G)_{ij}\bm{S}_i\cdot \bm{S}_j$. Therefore, according to the Luttinger-Tisza theory~\cite{bergman_order_2007,niggemann_classical_2019}, the classical ground state of $\mathcal{H}$ can be determined from the eigensolution of $A(G)$, of which the eigenvalues are defined as the \textit{spectrum} of the graph, denoted as $\sigma(G)$.

\begin{figure}[t!]
\includegraphics[width=0.48\textwidth]{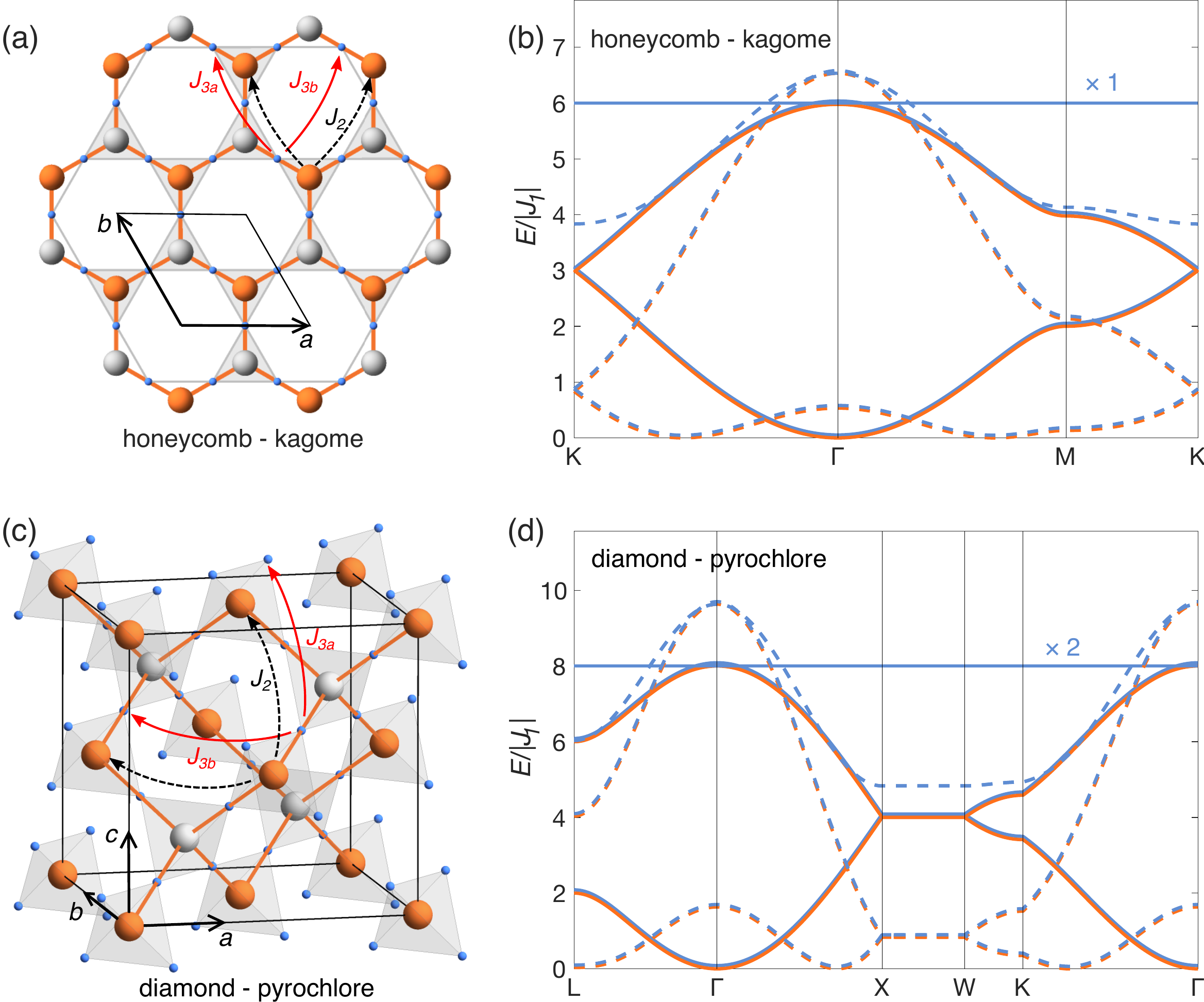}
\caption{(a) The line graph of a bipartite honeycomb lattice (large spheres linked by orange bonds) is a kagome lattice (small spheres linked by gray bonds). $J_1$ is the nearest-neighbor exchange coupling. Black dashed (red solid) arrows indicate the second-neighbor coupling $J_2$ (third-neighbor couplings $J_{3a}$ and $J_{3b}$) over the honeycomb (kagome) lattice. (b) Eigenbands of the interaction matrix of Heisenberg models on the honeycomb and kagome lattices. Orange solid (dashed) lines are the two eigenbands for the $J_1$ model ($J_1$-$J_2$ model with $J_2/J_1 = -0.25$) on the honeycomb lattice, which overlap with the two lower eigenbands of the $J_1$ model ($J_1$-$J_3$ model with $J_3/J_1 = -0.25$) on the kagome lattice shown in blue solid (dashed) lines. (c-d) Similar correspondence exists between the diamond and pyrochlore lattices. The interaction matrices for the $J_1$ model ($J_1$-$J_2$ model with $J_2/J_1 = -0.3$) on the diamond lattice and the $J_1$ model ($J_1$-$J_3$ model with $J_3/J_1 = -0.3$) on the pyrochlore lattice share two same eigenbands shown by the overlapping orange and blue solid (dashed) lines. 
\label{fig:band}}
\end{figure}

Algebraic graph theory defines that the related graphs should have related spectra~\cite{knauer_algebraic_2019, kollar_line_2020}. Of interest here is the line graph $L_G$~\cite{mielke_ferromagnetic_1991,tasaki_ferromagnetism_1992}, whose vertices correspond to the edges of the root graph $G$ and are adjacent if the original edges share a vertex: $L_G = (D, \{\{e,e'\}\,|\, e\,\cup\, e'\neq \emptyset,e \neq e'\})$. For a regular bipartite graph $G$, a convenient way to define its line graph is to select the vertices at the midpoints of the original edges. As illustrated in Fig.~\ref{fig:band}, for the honeycomb ($l = 3$) and diamond ($l = 4$) lattices that are the prototype hosts of the SSL, their line graphs form the kagome and pyrochlore lattices, respectively. According to graph theory~\cite{knauer_algebraic_2019, kollar_line_2020}, the spectra of $G$ and $L_G$ are related by
\begin{equation} \label{eq:spectrum}
\sigma(L_G) = (-2)^{m-n}\,\cup\,\sigma(G)+l-2\,\textrm{,}
\end{equation}
where $m$ ($n$) is the total number of edges (vertices) of the root graph $G$. In reciprocal space, Eq.~(\ref{eq:spectrum}) indicates the existence of flat eigenbands on $L_G$ with a degeneracy of $l-2$, which has been the focus of many recent studies~\cite{kollar_line_2020, ma_spin_2020, chiu_line_2021, nakai_perfect_2022}. More importantly, it reveals that the non-flat eigenbands of $\sigma(L_G)$ and $\sigma(G)$ share the same dispersion up to a constant of $l-2$, and their eigenvectors are related by the incident matrix as discussed in Ref.~\cite{mizoguchi_magnetic_2018}.

Such a spectrum correspondence can be immediately verified for Heisenberg models. On a regular lattice of $g$ sublattices and $N$ primitive cells, the interaction matrix of a $J_1$-only model
\begin{equation}
   \mathcal{J}^{\alpha\beta}_1(\bm{q})=\frac{J_1}{N}\sum_{\substack{i\in\alpha, j\in\beta \\ ij\in\langle ij\rangle}} \exp{[-i\bm{q}\cdot(\bm{r}_i - \bm{r}_j)]}
\end{equation}
is a $g\times g$ hollow matrix with zero diagonal elements. The eigenbands $\nu(\bm{q})$ for $J_1<0$ are shown as solid lines in Fig.~\ref{fig:band}(b) for the honeycomb and kagome lattices, and in Fig.~\ref{fig:band}(d) for the diamond and pyrochlore lattices. The minimum of the eigenbands $\nu_{\textrm{min}}$ has been subtracted for comparison. Analytical expressions for the eigenbands are presented in the Supplemental Material~\cite{supp}, which includes additional Refs.~\cite{buessen_spinmc_2020, hukushima_exchange_1996, cameron_magnetic_2016, izabela_study_2021, ye_implementation_2018, rodriguez_recent_1993, arnold_mantid_2014, farhi_ifit_2014, calder_suite_2018, toth_linear_2015, aroyo_crystal_2011}. Aside from the top flat bands in blue color, the two dispersive bands $\nu_{\pm}(\bm{q})-\nu_{\textrm{min}}$ on the kagome (pyrochlore) lattice overlap exactly with those on the honeycomb (diamond) lattice, which is a direct consequence of spectral graph theory.

This eigenband correspondence is maintained under the addition of certain further-neighbor interactions. For the $J_1$-$J_2$ model on a bipartite lattice, $J_2$ couples the spins of the same sublattices. Therefore, its contribution to $\mathcal{J}(\bm{q})$ is a diagonal matrix $\mathcal{J}_2(\bm{q})$ that commutes with $\mathcal{J}_1(\bm{q})$,  leading to a $\bm{q}$-dependent shift $\gamma_G(\bm{q})$ of the eigenbands with $\gamma_G(\bm{q})=(J_2/Ng)\sum_{ij\in\langle\!\langle ij\rangle\!\rangle} \exp{[-i\bm{q}\cdot(\bm{r}_i - \bm{r}_j)]}$, where $\langle\!\langle ij\rangle\!\rangle$ are the second-neighbor bonds~\cite{supp}. Similar conclusions can be drawn for the $J_1$-$J_3$ model on the line-graph lattices, as $J_3$ (including both $J_{3a}$ and $J_{3b}$) also couples spins of the same sublattices~\cite{supp}. Since $J_2(G)$ on the root-graph lattice and $J_3(L_G)$ on the line-graph lattice share the same exchange paths as compared in Fig.~\ref{fig:band}(a,c), we expect, in the case of $J_2(G)=J_3(L_G)$, the same dispersive eigenbands up to a constant on the root- and line-graph lattices. This correspondence is illustrated by the dashed lines in Fig.~\ref{fig:band}(b,d).

\textit{Approximate SSL.}--- An approximate SSL can be realized on the line-graph lattices through the eigenband correspondence. Following the results of the $J_1$-$J_2$ model on the bipartite lattices~\cite{bergman_order_2007,niggemann_classical_2019}, it is clear that for the $J_1$-$J_3$ model on a line-graph lattice the eigenband minima of $\mathcal{J}(\bm{q})$ will form a degenerate manifold in reciprocal space at $|J_3/J_1|>1/(2l)$, where $l$ is the number of the NN sites on the root-graph lattice. Here $J_1$ and $J_3$ should be FM and AF, respectively, as the additional flat bands on line graphs remove the FM-AF duality of a bipartite lattice. Since the equal moment constraint over the eigenvectors of $\mathcal{J}(\bm{q})$ is not always satisfied for $\bm{q}$ over the minimum manifold~\cite{supp}, the SSL realized through the line-graph approach is approximate and needs to be stabilized by thermal fluctuations. As discussed in the Supplemental Material~\cite{supp}, the degeneracy breaking over the spiral surface is relatively weak, which contrasts the strong modulation in the previously studied half-moon patterns~\cite{rau_spin_2016, mizoguchi_magnetic_2018, hering_fracton_2021, kiese_pinch_2022}. This weak degeneracy breaking on the line-graph lattices leads to a stable SSL state in a wide temperature regime that is comparable to that on the bipartite lattices.

\begin{figure}[t!]
   \includegraphics[width=0.48\textwidth]{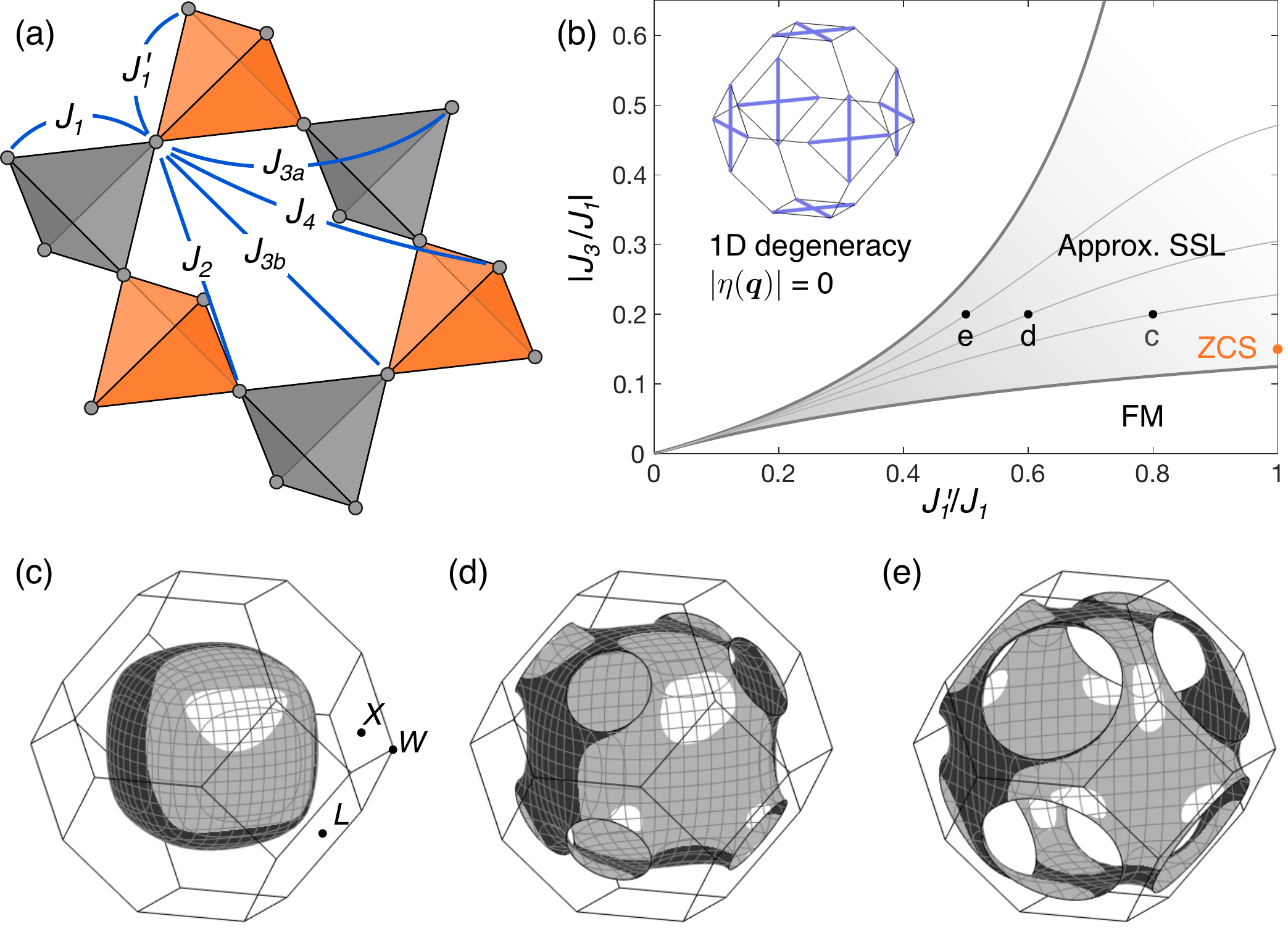}
   \caption{(a) The breathing pyrochlore lattice is composed of corner-sharing tetrahedra of two different sizes, which results in alternating $J_1$ and $J_1'$ couplings over the gray and orange tetrahedra, respectively. The exchange paths of the $J_2$, $J_{3a}$, $J_{3b}$, and $J_4$ interactions are also indicated. (b) Phase diagram on the breathing pyrochlore lattice with ferromagnetic $J_1$ and $J_1'$ interactions together with antiferromagnetic $J_3$ interactions assuming $J_{3a} = J_{3b}$. The shaded area indicates the region where an approximate SSL phase can be stabilized by thermal fluctuations. Contour lines for constant $|\eta(\bm{q})|$ are shown in the approximate SSL phase. Along these lines, the spiral surface stay the same. Location of ZnCr$_2$Se$_4$ (ZCS) with $J_1 = J_1'$ and $|J_3/J_1|=0.15$ is indicated by the yellow dot. Inset shows the degenerate manifold (blue lines) for $J_3>\frac{1}{4}\frac{J_1J_1'}{|J_1-J_1'|}$ out of the SSL regime. (c-e) Characteristic spiral surfaces with quasi-degenerate energies in the first Brillouin zone at $|J_3/J_1|=0.2$ and $J_1'/J_1=0.8$ (c), 0.6 (d), and 0.5  (e). These points are indicated on the phase diagram in (b) using the panel labels.
   \label{fig:breath}}
   \end{figure}

Before making comparisons to the experimental data, we further generalize the SSL model by incorporating a breathing distortion on the line-graph lattice so that the NN interactions are modulated alternately as $J_1$ and $J_1'$~\cite{hirschberger_skyrmion_2019, gao_order_2019, ghosh_breathing_2019}. For the breathing pyrochlore lattice shown in Fig.~\ref{fig:breath}(a), the eigenbands of $\mathcal{J}(\bm{q})$ can be solved as~\cite{supp}:
\begin{align}
   \nu_{1,2}&=J_3\kappa(\bm{q}) \pm \sqrt{4(J_1-J_1')^2+J_1J_1'|\eta(\bm{q})|^2} + (J_1+J_1')\textrm{,} \nonumber \\
   \nu_{3,4}&=J_3\kappa(\bm{q})-(J_1+J_1')\, \textrm{,} \nonumber
\end{align}
where $\eta(\bm{q}) = \sum_{n=1}^{4} \exp(-i\bm{q}\cdot\bm{d}_n)$ with $\bm{d}_n$ denoting the 4 bonding vectors around each spin site~\cite{supp} and $\kappa(\bm{q})=|\eta(\bm{q})|^2-4$. Assuming $J_1<J_1'<0$ and $J_3>0$, an approximate SSL state is realized for $\frac{1}{4}\frac{J_1J_1'}{|J_1+J_1'|}<J_3<\frac{1}{4}\frac{J_1J_1'}{|J_1-J_1'|}$. The corresponding phase diagram is presented in Fig.~\ref{fig:breath}(b). Representative spiral surfaces in the approximate SSL phase at $|J_3/J_1|=0.2$ are shown in Fig.~\ref{fig:breath}(c-e) for $J_1'/J_1 = 0.8$, 0.6, and 0.5, respectively. The surfaces are identical to those on the diamond lattice~\cite{bergman_order_2007}. Similar conclusions on the breathing kagome lattice are presented in the Supplemental Material~\cite{supp}.

\textit{ZnCr$_2$Se$_4$ with a regular pyrochlore lattice.}--- The Cr-based chalcogenide spinels present nearly ideal model compounds to demonstrate the proposed line-graph approach to SSLs. In these systems, $J_1$ is FM due to the 90$^{\circ}$ superexchange path, while $J_2\sim 0$ due to negligible orbital overlap~\cite{yaresko_electronic_2008,tymoshenko_pseudo_2017}. As the first example, we study the short-range spin correlations in ZnCr$_2$Se$_4$, where the Cr$^{3+}$ ($S=3/2$) ions form a regular pyrochlore lattice.

\begin{figure}[t!]
   \includegraphics[width=0.48\textwidth]{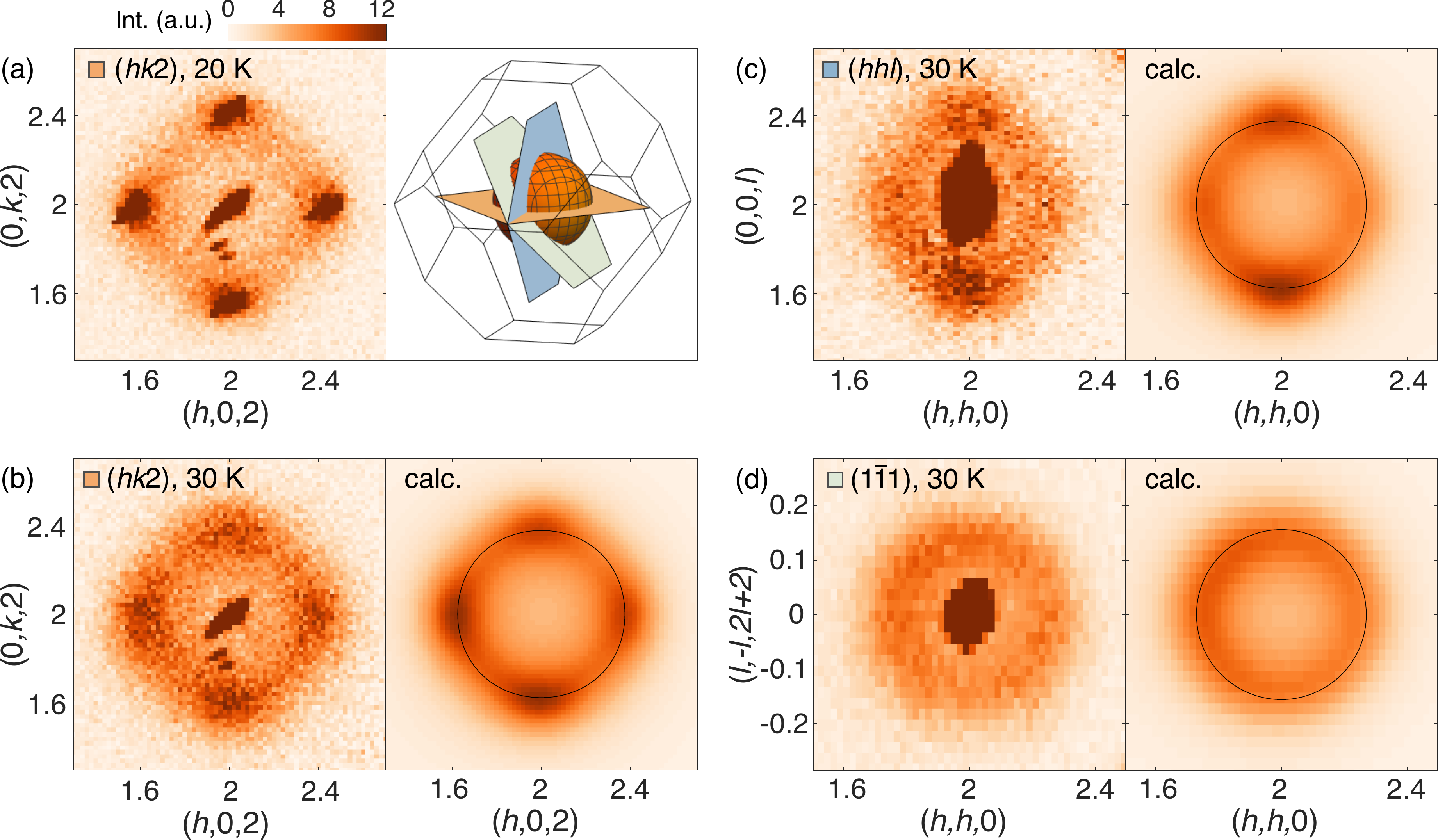}
   \caption{Quasistatic spin correlations in ZnCr$_2$Se$_4$ measured on CORELLI at  (a) 20~K and (b-d) 30~K. Slices are along the ($hk2$) plane in panels (a,b), ($hhl$) plane in panel (c), and (1$\overline{1}$1) plane in panel (d). The right sub-panel in (a) shows the complete spiral surface for the $J_1$-$J_3$ model on a pyrochlore lattice with $J_3/J_1= -0.15$. The flat planes correspond to the slice directions as indicated by the colored plaquette at the top left corner in each panel. The right sub-panels in (b-d) are the calculated diffuse scattering patterns for the fitted $J_1$-$J_2$-$J_3$-$J_4$ model. Solid line is the spiral surface predicted by the $J_1$-$J_3$ model with $J_3/J_1= -0.15$. The same linear intensity scale is employed in all panels.
   \label{fig:ZCSe}}
   \end{figure}

Single crystals of ZnCr$_2$Se$_4$ were grown using the chemical vapor transport method~\cite{supp}. Figure~\ref{fig:ZCSe} summarizes the diffuse neutron scattering results measured on CORELLI at the Spallation Neutron Source (SNS), Oak Ridge National Laboratory (ORNL)~\cite{supp}. With the statistical chopper, the elastic channel of our CORELLI data has an average energy resolution of about 0.8~meV for an incident neutron energy range of 13 to 33~meV. At 20~K, below the N\'eel transition temperature $T_N \sim 22$~K, magnetic Bragg peaks indexed by $\bm{q} = (0,0,0.47)$ are observed (see Fig.~\ref{fig:ZCSe}(a)). This is consistent with the helical ground state reported previously~\cite{hemberger_large_2007, yokaichiya_spin_2009, zajdel_structure_2017}, with the weak ring-like scattering mainly arising from the low-energy magnon excitations~\cite{tymoshenko_pseudo_2017,inosov_magnetic_2020}. At 30 K, above $T_N$, magnetic Bragg peaks are replaced by broad diffuse scattering with a spherical shape (see Fig.~\ref{fig:ZCSe}(b-d)), evidencing the emergence of an approximate SSL state where gapped excitations are replaced by quasielastic fluctuations~\cite{supp}.

   \begin{figure}[t!]
   \includegraphics[width=0.48\textwidth]{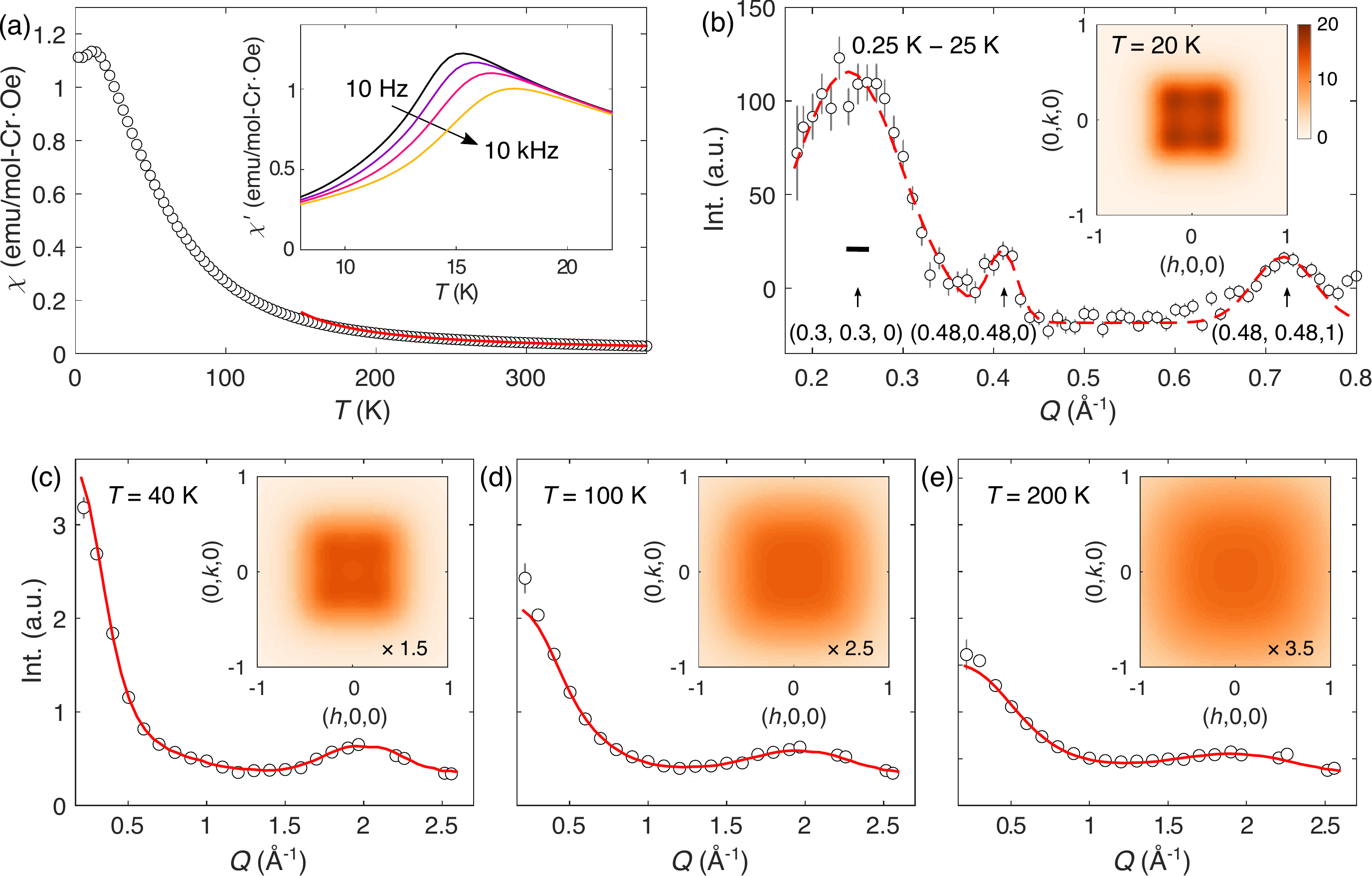}
   \caption{(a) DC magnetic susceptibility of CuInCr$_4$Se$_8$ measured in a $1\times10^4$~Oe field (circles). Red line shows the Curie Weiss fit at temperatures between 200 and 380~K. The fitted magnetic moment and Weiss temperature are $\mu_{\textrm{eff}}=4.6$~$\mu_B$/Cr and 97(2)~K, respectively. The inset shows the frequency dependence of the real part of the ac susceptibility measured in a 10~Oe field. Imaginary part of the ac susceptibility is presented in the Supplemental Material~\cite{supp}. (b) Difference between the neutron diffraction intensity at 0.25 and 25~K measured on HB-2A. The dashed line is a guide to the eyes. Positions of the characteristic wave vectors are indicated by arrows. The full width at half maximum (FHWM) of the closest nuclear reflection is indicated by the horizontal bar.  Inset shows the calculated neutron diffuse scattering pattern in the ($h,k,0$) plane at $T=20$~K for $J_{3a} = 0.07$~meV (see text). (c-e) Circles show the $Q$-dependence of the equal-time spin correlations at $T = 40$~(c), 100~(d), and 200~K~(e). Results of the SCGA fits using the $J_1$-$J_1'$-$J_{3a}$-$J_{3b}$ model are shown by the red solid lines. Insets are the calculated neutron diffuse scattering patterns in the ($h,k,0$) plane for $J_{3a} = 0.07$~meV at the corresponding temperatures (see text).
   \label{fig:CICSe}}
   \end{figure}

Assuming a Heisenberg model with exchange interactions up to the fourth neighbors ($J_4$), we fit the diffuse scattering data using the self consistent Gaussian approximation (SCGA) method~\cite{supp}. The calculated slices in Fig.~\ref{fig:ZCSe}(b-d) reproduce the experimental data. The coupling strengths are fitted as $J_1=-2.86(8)$, $J_2 = 0.00(1)$, $J_3 = 0.48(1)$, and $J_4 = -0.057(1)$~meV, which is very close to the values determined by inelastic neutron scattering (INS)~\cite{tymoshenko_pseudo_2017, inosov_magnetic_2020} and indicates marginal changes in the coupling strengths across the phase transition. The weak strengths of the $J_2$ and $J_4$ interactions allow a direct verification of the line-graph approach that is based on a $J_1$-$J_3$ model. The solid circles on top of the calculated patterns in Fig.~\ref{fig:ZCSe}(b-d) are the contours of the spiral surface predicted by the $J_1$-$J_3$ model with $J_3/J_1 = -0.15$. The contours capture the shapes of the diffuse scattering patterns. As compared in the Supplemental Material~\cite{supp}, the existence of the FM $J_4$ interaction slightly reduces the radius of the spiral surface while strongly modulates the scattering intensities.

\textit{CuInCr$_4$Se$_8$ with a breathing pyrochlore lattice.}--- Until now, we have assumed a uniform $J_3$ interaction. In real materials, the $J_3$ exchange paths on a line-graph lattice can be different as indicated in Fig.~\ref{fig:band}. This may lead to different coupling strengths and destabilize the SSL state. As the second example, we study the spin correlations in the breathing pyrochlore lattice compound CuInCr$_4$Se$_8$~\cite{duda_spin_2008}, which has been proposed as a SSL candidate from DFT calculations~\cite{ghosh_breathing_2019}.

A polycrystalline sample of CuInCr$_4$Se$_8$ was synthesized through the solid state reaction method~\cite{supp}. INS experiments were performed on SEQUOIA at the SNS, ORNL. Neutron diffraction experiments were performed on HB-2A at the high flux isotope reactor (HFIR), ORNL. As is consistent with the previous report~\cite{duda_spin_2008}, magnetic susceptibility shown in Fig.~\ref{fig:CICSe}(a) suggests a spin glass-like transition at $T_f \sim 15$~K with a clear frequency dependence. This is confirmed in the neutron diffraction results presented in Fig.~\ref{fig:CICSe}(b), where only broad magnetic features are observed down to 0.25~K. The weak features at $Q\sim0.41$ and 0.72~\AA$^{-1}$ can be indexed by $\bm{q}=(0.48, 0.48, 0)$, while their intensities exhibit a history dependence as expected for a spin glass state.

Figures~\ref{fig:CICSe}(c-e) present the equal-time spin correlations obtained by integrating the INS spectra from [$-20$, 20]~meV~\cite{supp}. Assuming a $J_1$-$J_1'$-$J_{3a}$-$J_{3b}$ Heisenberg spin model, we fit the diffuse scattering data by the SCGA method~\cite{supp, pokharel_cluster_2020}. The fitted results are shown in Fig.~\ref{fig:CICSe}(c-d) as solid lines. The fitted coupling strengths are $J_1=-1.6(2)$, $J_1' = -5.4(3)$, $J_{3a}=0.1(1)$, and $J_{3b}=0.6(1)$~meV, where the different strengths for $J_{3a}$ and $J_{3b}$ are necessary for a satisfactory fit. As shown in the inset of Fig.~\ref{fig:CICSe}(b-e), the spin correlations in the ($hk0$) plane does not follow a circular shape, implying the absence of a SSL state due to the unequal $J_{3a}$ and $J_{3b}$ interactions.

The fitted parameter set provides an explanation for the glassy ground state in CuInCr$_4$Se$_8$. $J_{3a}$, being the weakest interaction in the $J_1$-$J_1'$-$J_{3a}$-$J_{3b}$ model, is however the most crucial parameter in determining the exact length of the LRO $\bm{q}=(q,q,0)$. Within the standard deviation of $J_{3a}$, $q$ varies from 0 at $J_{3a}=0$ up to $\sim0.4$ at $J_{3a}=0.2$~meV. Therefore, the broad diffuse scattering at $Q\sim 0.25$~\AA$^{-1}$, which is tentatively indexed as $(0.3, 0.3, 0)$ in Fig.~\ref{fig:CICSe}(b), may arise from a finite distribution in $J_{3a}$ due to a tiny amount of structural defects. Weaker features indexed by $\bm{q}=(0.48, 0.48, 0)$ may be stabilized by local structural distortions considering its proximity to the commensurate $(\frac{1}{2},\frac{1}{2}, 0)$ position~\cite{lee_neel_2008, gao_manifolds_2018}. Studies of a single crystal sample will help verify the proposed mechanism.

\textit{Conclusion.}--- We have presented a line-graph approach to achieve approximate SSLs on non-bipartite lattices, which allows the experimental exploration of SSLs in a broader family of compounds. 
Besides ZnCr$_2$Se$_4$ studied in this work, chalcogenide spinels like ZnCr$_2$S$_4$~\cite{yokaichiya_spin_2009} and HgCr$_2$S$_4$~\cite{tsurkan_experimental_2006} are worth further investigations as an incommensurate helical ground state has been observed. The approximate SSLs on the line-graph lattices also provide a new platform to explore field-induced topological spin textures since the helicity of the single-$\bm{q}$ component is maintained~\cite{gao_fractional_2020}. Such a mechanism may account for the magnetic skyrmions in the breathing kagome-lattice compound Gd$_3$Ru$_4$Al$_{12}$~\cite{hirschberger_skyrmion_2019}. 
On the theoretical side, it remains an open question whether the approximate SSLs may evolve into quantum spin liquids under sufficient quantum fluctuations. It is also interesting to explore whether the fracton physics predicted for the exact SSLs~\cite{yan_low_2021} may survive to some extent in the approximate SSLs.

\begin{acknowledgments}
We acknowledge helpful discussions with Jyong-Hao Chen. This work was supported by the U.S. Department of Energy, Office of Science, Basic Energy Sciences, Materials Sciences and Engineering Division. This research used resources at the Spallation Neutron Source (SNS) and the High Flux Isotope Reactor (HFIR), both are DOE Office of Science User Facilities operated by the Oak Ridge National Laboratory (ORNL). G.P. acknowledges support from the Gordon and Betty Moore Foundation's  EPiQS Initiative, Grant GBMF9069.
\end{acknowledgments}


%

   \clearpage
   \newpage
   
   \renewcommand{\thefigure}{S\arabic{figure}}
   \renewcommand{\thetable}{S\arabic{table}}
   \renewcommand{\theequation}{S\arabic{equation}}

   \makeatletter
   \renewcommand*{\citenumfont}[1]{S#1}
   \renewcommand*{\bibnumfmt}[1]{[S#1]}
   \def\clearfmfn{\let\@FMN@list\@empty}  
   \makeatother
   \clearfmfn

   \setcounter{figure}{0} 
   \setcounter{table}{0}
   \setcounter{equation}{0} 
   
   \onecolumngrid
   \begin{center} {\bf \large Supplemental Materials for `Line-Graph Approach to Spiral Spin Liquids'} \end{center}
   
   \vspace{0.5cm}

   \section{Numerical solution of the interaction matrix}
As discussed in the main text, the spectrum correspondence between the root and line graphs can be directly verified for Heisenberg spin models. In this section, we present the analytical expressions for the interaction matrices and eigenvalues of the representative models.

\begin{figure}[b!]
   \includegraphics[width=0.7\textwidth]{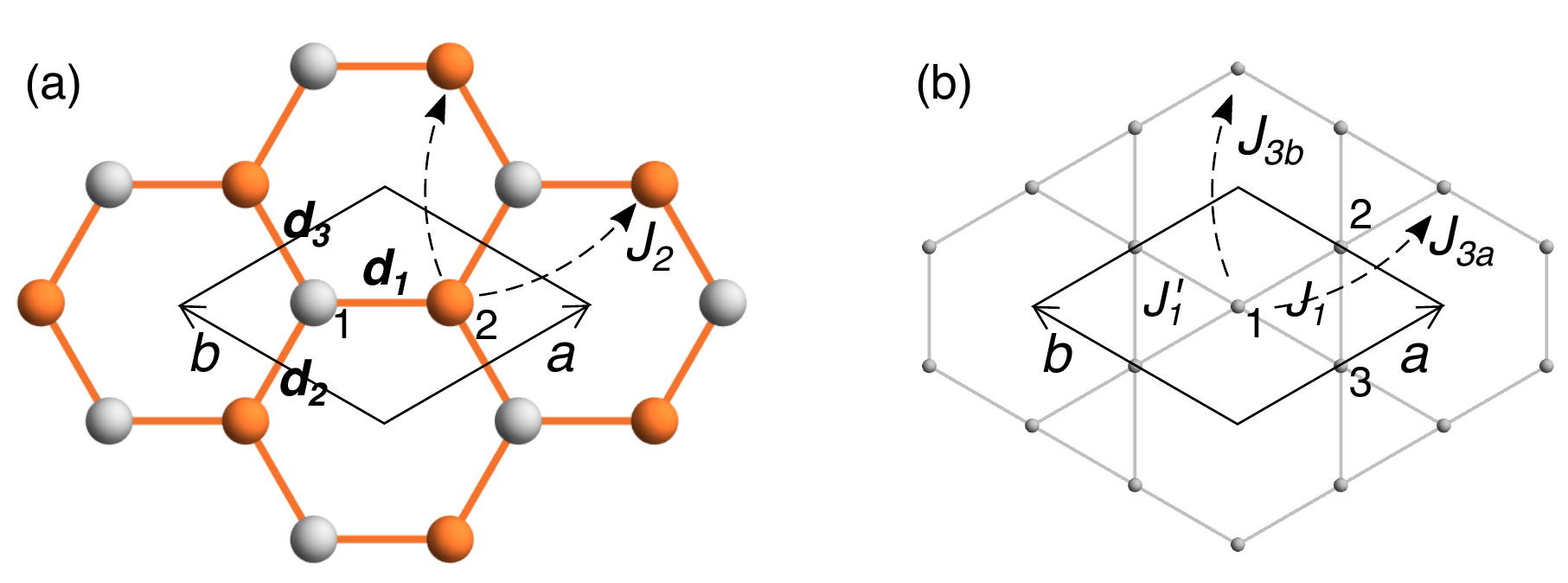}
   \caption{The honeycomb (a) and kagome (b) lattices with the sublattices indicated. The bonding vectors $\bm{d}_1$, $\bm{d}_2$, and $\bm{d}_3$ on the honeycomb lattice are shown in panel (a). Exchange paths for the $J_2$ ($J_{3a}$ and $J_{3b}$) interactions on the honeycomb (kagome) lattice are shown as dashed lines.  In the presence of a breathing distortion, the exchange interactions over the neighboring triangles on the kagome lattice will have two different strengths of $J_1$ and $J_1'$ as indicated in panel (b).
   \label{fig:honey}}
   \end{figure}

\subsection{\texorpdfstring{$\boldsymbol{J_1-J_2}$}{pdf} model on a honeycomb lattice}
The $J_1$-$J_2$ Heisenberg model on a bipartite honeycomb lattice (see Fig.~\ref{fig:honey}(a)) is a well-established model that hosts SSL~\cite{mulder_spiral_2010s,niggemann_classical_2019s}. Here we present the main results for comparison. Following the definition in the main text, the interaction matrix $\mathcal{J}(\bm{q})$ of the spin Hamiltonian $\mathcal{H} = J_1\sum_{\langle ij\rangle}\bm{S}_i\cdot\bm{S}_j + J_2\sum_{\langle\langle ij\rangle\rangle}\bm{S}_i\cdot\bm{S}_j$ on a honeycomb lattice can be expressed as 
\begin{align}\label{eq:honey}
   \mathcal{J}(\bm{q})=J_1\begin{pmatrix} 0 & \zeta \\ \zeta^* & 0 \end{pmatrix} + J_2\gamma\mathbb{1} \text{,}
\end{align}
where the off-diagonal variable is $\zeta(\bm{q}) = \sum_{n=1}^{3} \exp(-i\bm{q}\cdot\bm{d}_n)$ with bonding vectors $\bm{d}_1 = \frac{1}{3}(\bm{b}-\bm{a})$, $\bm{d}_2 = \bm{d}_1 + \bm{a}$, and $\bm{d}_3 = \bm{d}_1-\bm{b}$ shown in Fig.~\ref{fig:honey}(a). The diagonal variable is $\gamma(\bm{q})=|\zeta(\bm{q})|^2-3$. The eigenvalues of $\mathcal{J}(\bm{q})$ can be solved as 
\begin{equation}
   \nu_{\pm}(\bm{q})=J_2\gamma(\bm{q}) \pm |J_1\zeta(\bm{q})|. 
\end{equation}
A SSL is realized in the regime of $|J_2/J_1|> 1/6$. As discussed in Ref.~\cite{niggemann_classical_2019s}, the eigenvectors of $\mathcal{J}(\bm{q})$ have equal components over the two sublattices, which indicates an exact degeneracy over the spiral surface according to the Luttinger-Tisza theory.

\subsection{\texorpdfstring{$\boldsymbol{J_1-J_3}$}{pdf} model on a kagome lattice}
For the $J_1$-$J_3$ Heisenberg model on a kagome lattice (see Fig.~\ref{fig:honey}(b)), the interaction matrix is
\begin{align}
   \mathcal{J}(\bm{q})=J_1\begin{pmatrix} 0 & \xi_{12}& \xi_{13} \\ \xi_{12} & 0 & \xi_{23} \\ \xi_{13} & \xi_{23} & 0 \end{pmatrix} + J_3\gamma\mathbb{1} \text{,}
\end{align}
where the off-diagonal variables are $\xi_{12}(\bm{q})=2\cos(\pi q_h)$, $\xi_{13}(\bm{q})=2\cos(\pi q_k)$, and $\xi_{13}(\bm{q})=2\cos(\pi q_h + \pi q_k)$. The diagonal variable $\gamma(\bm{q})$ is the same as that in Eq.~(\ref{eq:honey}) for the honeycomb lattice. The eigenvalues can be solved as 
\begin{align}
&\nu_{1,2} = J_3\gamma(\bm{q}) \pm |J_1\zeta(\bm{q})|+J_1, \\
&\nu_{3} = J_3\gamma(\bm{q})-2J_1.
\end{align}
Here $\zeta(\bm{q})$ is the same as that defined in Eq.~(\ref{eq:honey}). It is immediately clear that when $J_3$ on the kagome lattice equals $J_2$ on the honeycomb lattice, the dispersion of $\nu_{1,2}$  is exactly the same as $\nu_{\pm}$ on the honeycomb lattice up to a constant of $J_1$, which is a direct consequence of the spectral graph theory.

In spite of the similar eigenvalues, it can be shown numerically that the eigenvectors of $\mathcal{J}(\bm{q})$ on the kagome lattice generally have close but unequal components over the three sublattices, which contrasts the equal component eigenvectors on the honeycomb lattice. This means the degeneracy over the spiral surface on the the kagome lattice is not exact, leading to an approximate SSL state that is only stabilized by thermal fluctuations. 

\begin{figure}[t!]
   \includegraphics[width=0.9\textwidth]{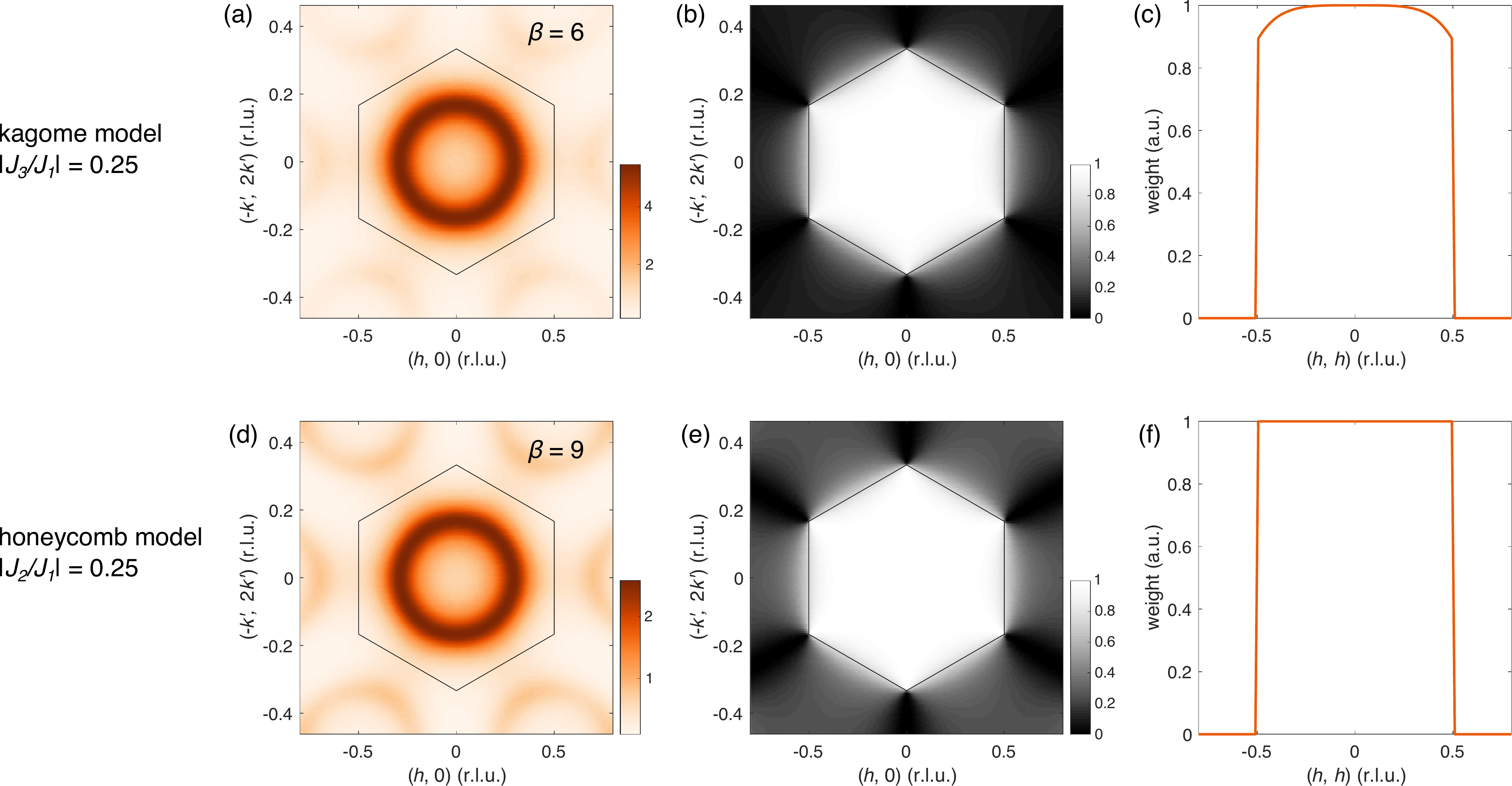}
   \caption{(a) Spin correlation pattern of the kagome-lattice model with $J_1 = -1$ and $J_3 = 0.25$ at inverse temperature $\beta= 6$. The first Brillouin zone is outlined by black lines. Intensities are shown in linear scale. (b) Pseudocolor map of the weight factor calculated from the eigenvectors of the kagome model. (c) Weight factor along the $(h,h)$ direction. (d-f) Similar plots as (a-c) for the honeycomb-lattice model with $J_1 = -1$ and $J_2 = 0.25$.
   \label{fig:weight}}
   \end{figure}

The approximate nature of the SSL state on the kagome lattice can also be seen from the $\bm{q}$-dependence of the weight factor $\sum_{\mu,\nu}[\psi^*(\bm{q})]_\mu[\psi(\bm{q})]_\nu$, where $\psi$ is the eigenfunction for the lowest eigenband of the interaction matrix~\cite{mizoguchi_magnetic_2018s}. Figure~\ref{fig:weight} compares the weight factor for the kagome- and honeycomb-lattice models with the same frustration ratio of 0.25. Like the diffuse scattering patterns calculated by the SCGA method shown in Fig.~\ref{fig:weight}(a) and (c), the weight factor for on the kagome and honeycomb lattices shown in Fig.~\ref{fig:weight}(b) and (e) are very similar to each other, thus justifying the existence of a SSL state in the former model. However, as compared in Fig.~\ref{fig:weight}(c) and (f) for line cuts along the $(h,h)$ direction, the weight factor of the kagome model exhibits an additional $\bm{q}$-dependence near the Brillouin zone boundary. Such a $\bm{q}$-dependence contrasts the flatter weight factor of the honeycomb model and indicates the approximate nature of the SSL state on the kagome lattice.

\begin{figure}[b!]
   \includegraphics[width=0.5\textwidth]{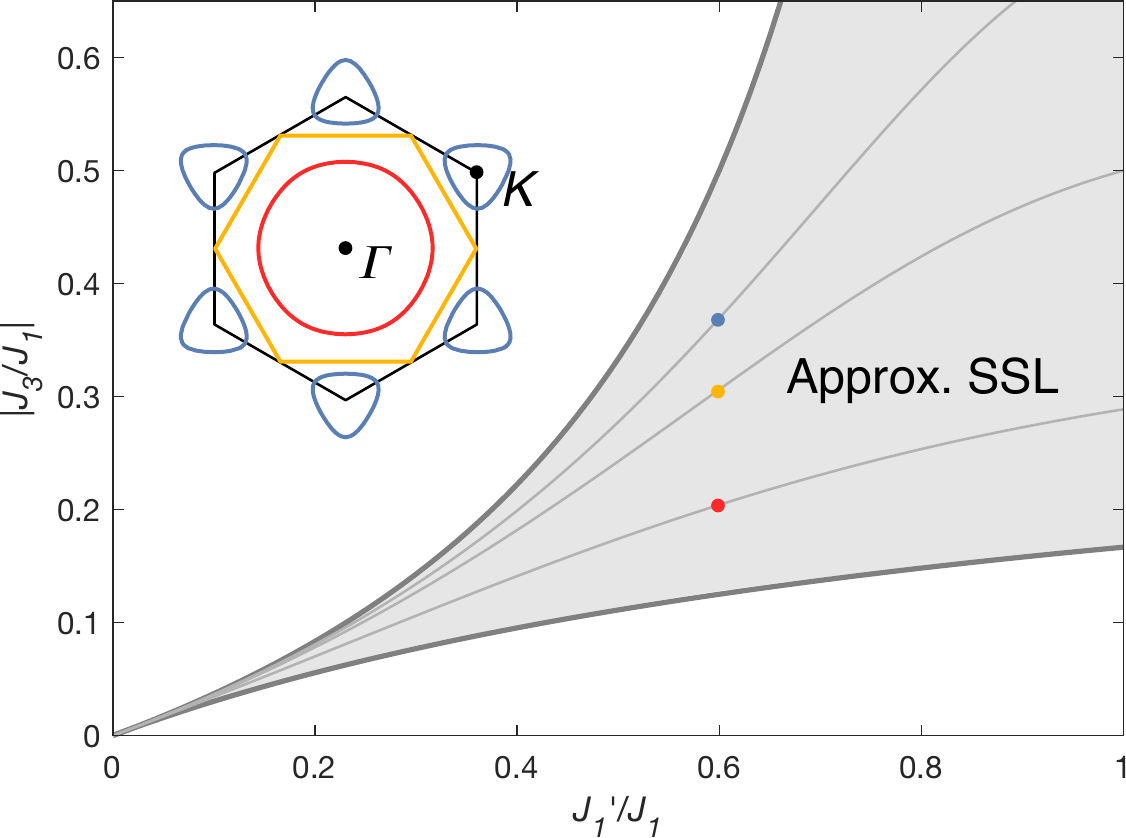}
   \caption{Phase diagram on the breathing kagome lattice with ferromagnetic $J_1$ and $J_1'$ interactions together with antiferromagnetic $J_3$. The shaded area indicates the region of the approximate SSL phase. Contour lines for constant $|\zeta(\bm{q})|$ are shown in the approximate SSL phase. Inset shows the representative spiral surface in reciprocal space at $|J_3/J_1|=0.204$ (red), 0.306 (yellow), and 0.369 (blue) with $J_1'$ fixed at $0.6J_1$, which correspond to $|\zeta(\bm{q})|^2=3$, 1, and 0.5, respectively.
   \label{fig:pkago}}
   \end{figure}

\subsection{\texorpdfstring{$\boldsymbol{J_1-J_1'-J_3}$}{pdf} model on a breathing kagome lattice}
On the breathing kagome lattice, the hollow matrix term in the interaction matrix needs to be revised to incorporate the different coupling strengths $J_1$ and $J_1'$ over the neighboring triangles. The interaction matrix of the $J_1$-$J_1'$-$J_3$ Heisenberg model on a breathing kagome lattice can be expressed as
\begin{align}
   \mathcal{J}(\bm{q})=J_1\begin{pmatrix} 0 & \varepsilon_{12}& \varepsilon_{13} \\ \varepsilon_{12}^* & 0 & \varepsilon_{23} \\ \varepsilon_{13}^* & \varepsilon_{23}^* & 0 \end{pmatrix} + J_3\gamma\mathbb{1} \text{,}
\end{align}
where
\begin{align}
 \varepsilon_{12}(\bm{q})&=J_1\exp(i\pi q_h)+ J_1'\exp(-i\pi q_h) ,\\ 
 \varepsilon_{13}(\bm{q})&=J_1\exp(-i\pi q_k)+ J_1'\exp(i\pi q_k),\\
 \varepsilon_{23}(\bm{q})&=J_1\exp(-i\pi q_h-i\pi q_k)+ J_1'\exp(i\pi q_h + i\pi q_k).
\end{align}
The definition of $\gamma(\bm{q})$ is the same as that in Eq.~(\ref{eq:honey}). The eigenvalues can be solved as
\begin{align}
   &\nu_{1,2} = J_3\gamma(\bm{q}) \pm \sqrt{\frac{9(J_1-J_1')^2}{4}+J_1J_1'|\zeta(\bm{q})|^2}+\frac{J_1+J_1'}{2}, \\
   &\nu_{3} = J_3\gamma(\bm{q})-(J_1+J_1').
\end{align}
Assuming $J_1<J_1'<0$ and $J_3>0$, it can be shown that an approximate SSL phase is stabilized in the parameter regime of
\begin{align}
   \frac{1}{3}\frac{J_1J_1'}{|J_1+J_1'|}<J_3<\frac{1}{3}\frac{J_1J_1'}{|J_1-J_1'|},
\end{align}
and the phase diagram is summarized in Fig.~\ref{fig:pkago}. the eigenvalue minimum exhibits a continuous degeneracy. The spiral surface is defined by
\begin{align}
   |\zeta(\bm{q})|^2= \frac{J_1J_1'}{4J_3^2}-\frac{9}{4}\frac{(J_1-J_1')^2}{J_1J_1'}.
\end{align}
Typical spiral surfaces at $|J_3/J_1|=0.204$, 0.306, and 0.369 with $J_1'$ fixed at $0.6J_1$ are shown in the inset of Fig.~\ref{fig:pkago}. Similar to the results on a reagular kagome lattice, this SSL has approximate degeneracy.

\subsection{\texorpdfstring{$\boldsymbol{J_1-J_2}$}{pdf} model on a diamond lattice}

For the $J_1$-$J_2$ Heisenberg model on a bipartite diamond lattice (see Fig.~\ref{fig:diamond}(a)), the interaction matrix is
\begin{align}\label{eq:diamond}
   \mathcal{J}(\bm{q})=J_1\begin{pmatrix} 0 & \eta \\ \eta^* & 0 \end{pmatrix} + J_2\kappa\mathbb{1} \text{,}
\end{align}
where the off-diagonal variable is $\eta(\bm{q}) = \sum_{n=1}^{4} \exp(-i\bm{q}\cdot\bm{d}_n)$ with bonding vectors $\bm{d}_1 = \frac{1}{4}(\bm{a}+\bm{b}+\bm{c})$, $\bm{d}_2 = \frac{1}{4}(\bm{a} - \bm{b} -\bm{c})$, $\bm{d}_3 = \frac{1}{4}(-\bm{a}+\bm{b}-\bm{c})$, and $\bm{d}_4 = \frac{1}{4}(-\bm{a}-\bm{b}+\bm{c})$ as indicated in Fig.~\ref{fig:diamond}(a). The diagonal variable is $\kappa(\bm{q})=|\eta(\bm{q})|^2-4$. The eigenvalues can be solved as
\begin{align}
\nu_{\pm}= J_2\kappa({\bm{q}})\pm|J_1\eta(\bm{q})|.
\end{align} 
It has been established that a SSL state with exact degeneracy emerges in the regime of $|J_2/J_1|>1/8$~\cite{bergman_order_2007s}.

\subsection{\texorpdfstring{$\boldsymbol{J_1-J_3}$}{pdf} model on a pyrochlore lattice}
\begin{figure}[t!]
   \includegraphics[width=0.7\textwidth]{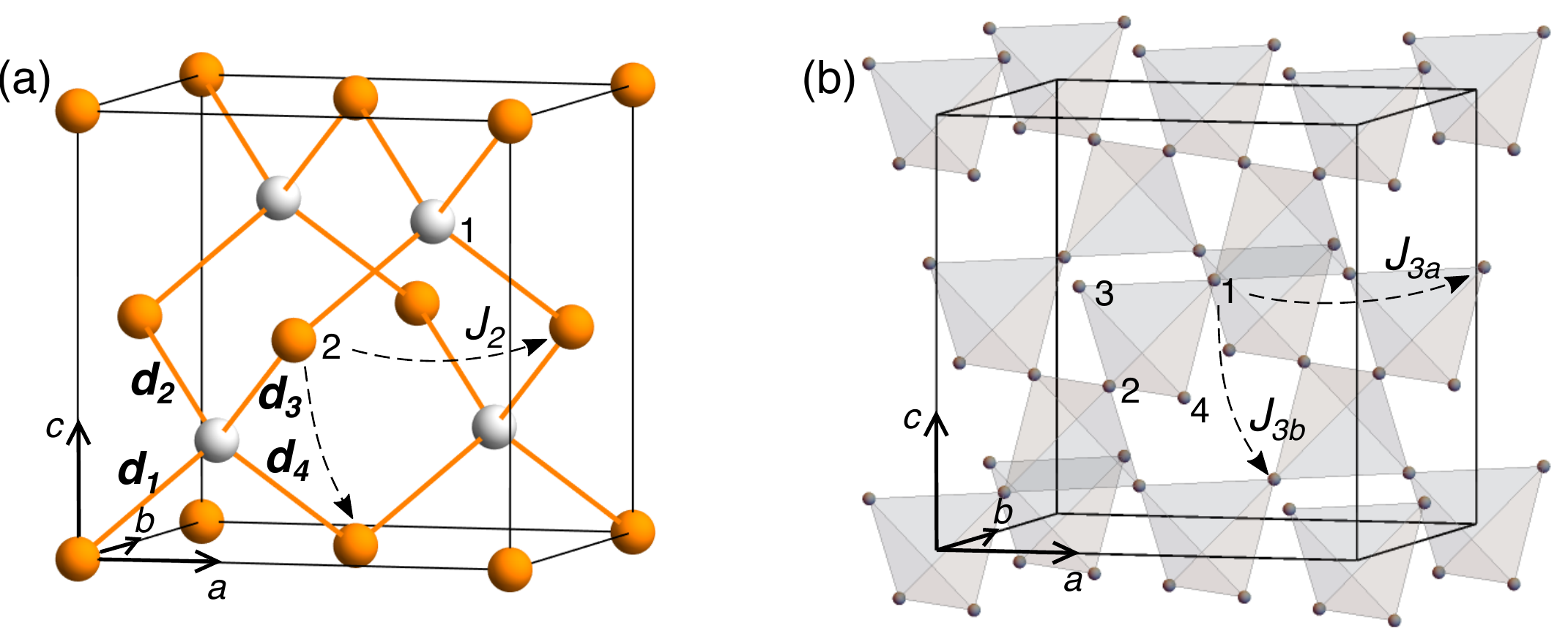}
   \caption{The diamond (a) and pyrochlore (b) lattices with the sublattices indicated. The bonding vectors $\bm{d}_1$, $\bm{d}_2$, $\bm{d}_3$, and $\bm{d}_4$ on the diamond lattice are shown in panel (a). Exchange paths for the $J_2$ ($J_{3a}$ and $J_{3b}$) interactions on the diamond (pyrochlore) lattice are shown as dashed lines. In the presence of a breathing distortion, the exchange interactions over the neighboring tetrahedra on the pyrochlore lattice will have two different strengths of $J_1$ and $J_1'$.
   \label{fig:diamond}}
   \end{figure}

For the $J_1$-$J_3$ Heisenberg model on a pyrochlore lattice (see Fig.~\ref{fig:diamond}(b)), the interaction matrix is
\begin{align}
   \mathcal{J}(\bm{q})=J_1\begin{pmatrix} 0 & \varrho_{12}& \varrho_{13} & \varrho_{14} \\ \varrho_{12} & 0 & \varrho_{23} & \varrho_{24} \\ \varrho_{13} & \varrho_{23} & 0 & \varrho_{34} \\ \varrho_{14} & \varrho_{24} &  \varrho_{34} & 0 \end{pmatrix} + J_3\kappa\mathbb{1} \text{,}
\end{align}
where the off-diagonal variables are $\varrho_{12}(\bm{q})=2\cos(\frac{\pi q_h + \pi q_l}{2})$, $\varrho_{13}(\bm{q})=2\cos(\frac{\pi q_h + \pi q_k}{2})$, $\varrho_{14}(\bm{q})=2\cos(\frac{\pi q_k + \pi q_l}{2})$, $\varrho_{23}(\bm{q})=2\cos(\frac{\pi q_k - \pi q_l}{2})$, $\varrho_{24}(\bm{q})=2\cos(\frac{\pi q_h - \pi q_k}{2})$, $\varrho_{34}(\bm{q})=2\cos(\frac{\pi q_h - \pi q_l}{2})$. The diagonal variable $\kappa(\bm{q})$ is the same as that in Eq.~(\ref{eq:diamond}) for the diamond lattice.  The eigenvalues of $\mathcal{J}(\bm{q})$ can be solved as 
\begin{align}
\nu_{1,2} &=J_3\kappa({\bm{q}})\pm|J_1\eta(\bm{q})| + 2J_1,\\
\nu_{3,4} &=J_3\kappa({\bm{q}})-2J_1.
\end{align}

Similar to the results on the kagome lattice, numerical solutions show that the eigenvectors generally have unequal components over the four sublattices. Therefore, the SSL on the pyrochlore lattice has approximate degeneracy rather than exact degeneracy as that on the diamond lattice.

\subsection{\texorpdfstring{$\boldsymbol{J_1-J_1'-J_3}$}{pdf} model on a breathing pyrochlore lattice}
On a breathing pyrochlore lattice, the hollow matrix term in the interaction matrix needs to be revised to incorporate the different coupling strengths $J_1$ and $J_1'$ over the neighboring tetrahedra. Defining
\begin{align}
   \tau_{ab}(\bm{q}) &= J_1\exp(\pi q_a + \pi q_b) + J_1'\exp(-\pi q_a - \pi q_b)\\
   \tau_{\overline{ab}}(\bm{q}) &= J_1\exp(\pi q_a - \pi q_b) + J_1'\exp(-\pi q_a + \pi q_b)\textrm{,}
\end{align}
the interaction matrix of a $J_1$-$J_1'$-$J_3$ Heisenberg model on a breathing pyrochlore lattice can be expressed as
\begin{align}
   \mathcal{J}(\bm{q})=J_1\begin{pmatrix} 0 & \tau_{hl}& \tau_{hk} & \tau_{kl} \\
       \tau_{hl}^* & 0 & \tau_{\overline{kl}} & \tau_{\overline{hk}}^* \\ 
       \tau_{hk}^* & \tau_{\overline{kl}}^* & 0 & \tau_{\overline{hl}}^* \\ 
       \tau_{kl}^* & \tau_{\overline{hk}} &  \tau_{\overline{hl}} & 0 \end{pmatrix}
        + J_3\kappa\mathbb{1} \text{,}
\end{align}
where $\kappa(\bm{q})$ is the same as that in Eq.~(\ref{eq:diamond}) for the diamond lattice. The eigenvalues can be solved as 
\begin{align}
\nu_{1,2}&=J_3\kappa(\bm{q}) \pm \sqrt{4(J_1-J_1')^2+J_1J_1'|\eta(\bm{q})|^2} + (J_1+J_1') \\
\nu_{3,4}&=J_3\kappa(\bm{q})-(J_1+J_1').
\end{align}
Here $\eta(\bm{q})$ is the same as that defined in Eq.~(\ref{eq:diamond}). Assuming $J_1<J_1'<0$ and $J_3>0$, an approximate SSL phase is realized in the parameter regime of 
\begin{align}
   \frac{1}{4}\frac{J_1J_1'}{|J_1+J_1'|}<J_3<\frac{1}{4}\frac{J_1J_1'}{|J_1-J_1'|}.
\end{align}
The spiral surface in reciprocal space is defined by
\begin{align}
   |\eta(\bm{q})|^2= \frac{J_1J_1'}{4J_3^2}-\frac{4(J_1-J_1')^2}{J_1J_1'}.
\end{align}
The phase diagram and representative spiral surfaces are presented in Fig.~2 of the main text. As on the regular pyrochlore lattice, the degeneracy in this SSL state is approximate.

\begin{figure}[b!]
   \includegraphics[width=1.0\textwidth]{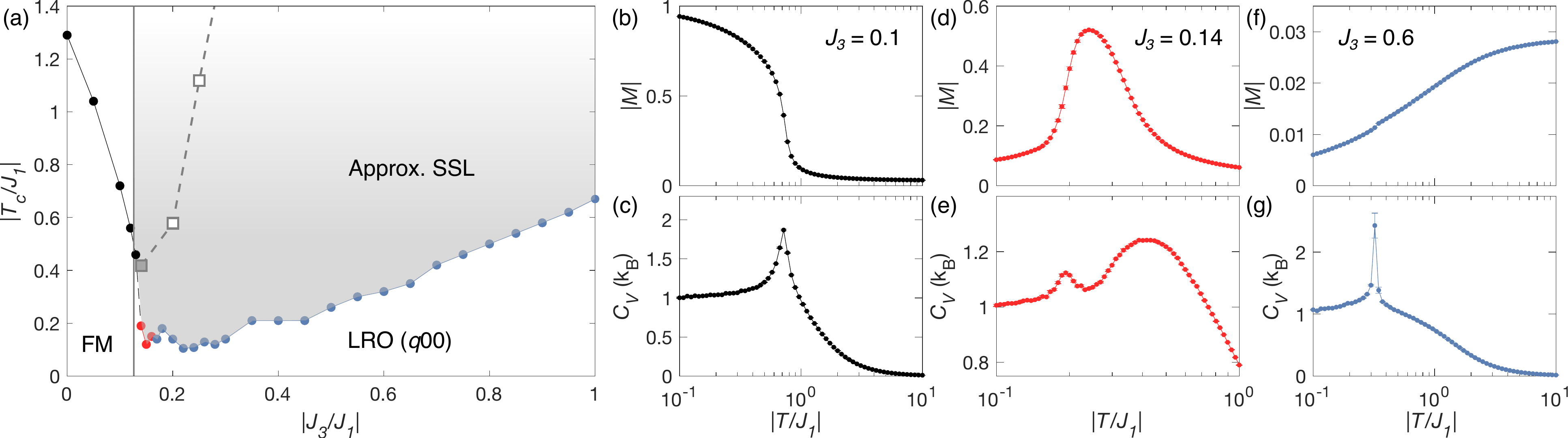}
   \caption{(a) Circular markers are the critical temperature of the $J_1$-$J_3$ model on a regular pyrochlore lattice as a function of $|J_3/J_1|$. $J_1$ and $J_3$ are assumed to be FM and AFM, respectively. Empty (solid) squares indicate the crossover transition between the paramagnetic and SSL states obtained from the maximum of the broad hump in magnetization for $J_3 = 0.2$ and 0.25 (heat capacity for $J_3 = 0.14$). The broad hump in magnetization for $J_3 = 0.14$ is arising from a precursory FM phase instead of the crossover transition. (b-g) Temperature dependence of the magnetization ($|M|$) and heat capacity ($C_V$) per spin for $|J_3/J_1|=0.1$ (b, c), 0.14 (d, e), and 0.6 (f, g). These three parameter sets are representative for the regions shown in black, red, and blue over the phase diagram in panel (a).
   \label{fig:montecarlo}}
   \end{figure}

\section{Stability of the approximate SSL phase}
Although the degeneracy of the SSL state on a non-bipartite lattice is only approximate, the energy difference among the spiral configurations over the spiral surface is small, thus ensuring a robust SSL state down to low temperatures. As an illustration, we perform classical Monte Carlo simulations for the $J_1$-$J_3$ model on a pyrochlore lattice. Our simulations are implemented with the SpinMC.jl package~\cite{buessen_spinmc_2020s} that employs a parallel tempering scheme~\cite{hukushima_exchange_1996s} to facilitate thermal equilibration. Figure~\ref{fig:montecarlo} summarizes the LRO transition temperature $|T_c/J_1|$ as a function of $|J_3/J_1|$, which is calculated on a supercell of $16\times4^3$ spins with up to $5\times10^5$ thermalization sweeps and $4\times10^6$ measurement sweeps. Three characteristic regimes are observed from the temperature dependence of the magnetization ($|M|$) and heat capacity ($C_V$) per spins as summarized in Fig.~\ref{fig:montecarlo}(b-g): At $|J_3/J_1|<=0.13$ (black), the system exhibits a FM ground state; At $|J_3/J_1|>=0.17$ (blue), the system exhibits a AFM ground state with propagation vector along the (100) direction; A transitional regime with strong FM fluctuation is observed at $0.13<|J_3/J_1|<0.17$ (red). 

The most important observation in our Monte Carlo simulations is that the transition temperature $T_c$ in the SSL regime is rather low. Comparing to the $J_1$-$J_2$ model on a diamond lattice where LRO is induced by an order-by-disorder transition~\cite{bergman_order_2007s}, $T_c$ of the $J_1$-$J_3$ model on a pyrochlore lattice is only $\sim2$ times higher. Considering the difference in the number of nearest-neighbor bonds, which is 4 and 6 on the diamond and pyrochlore lattices, respectively, we expect very similar stability regime of the SSL states.

\begin{figure}[h!]
   \includegraphics[width=0.9\textwidth]{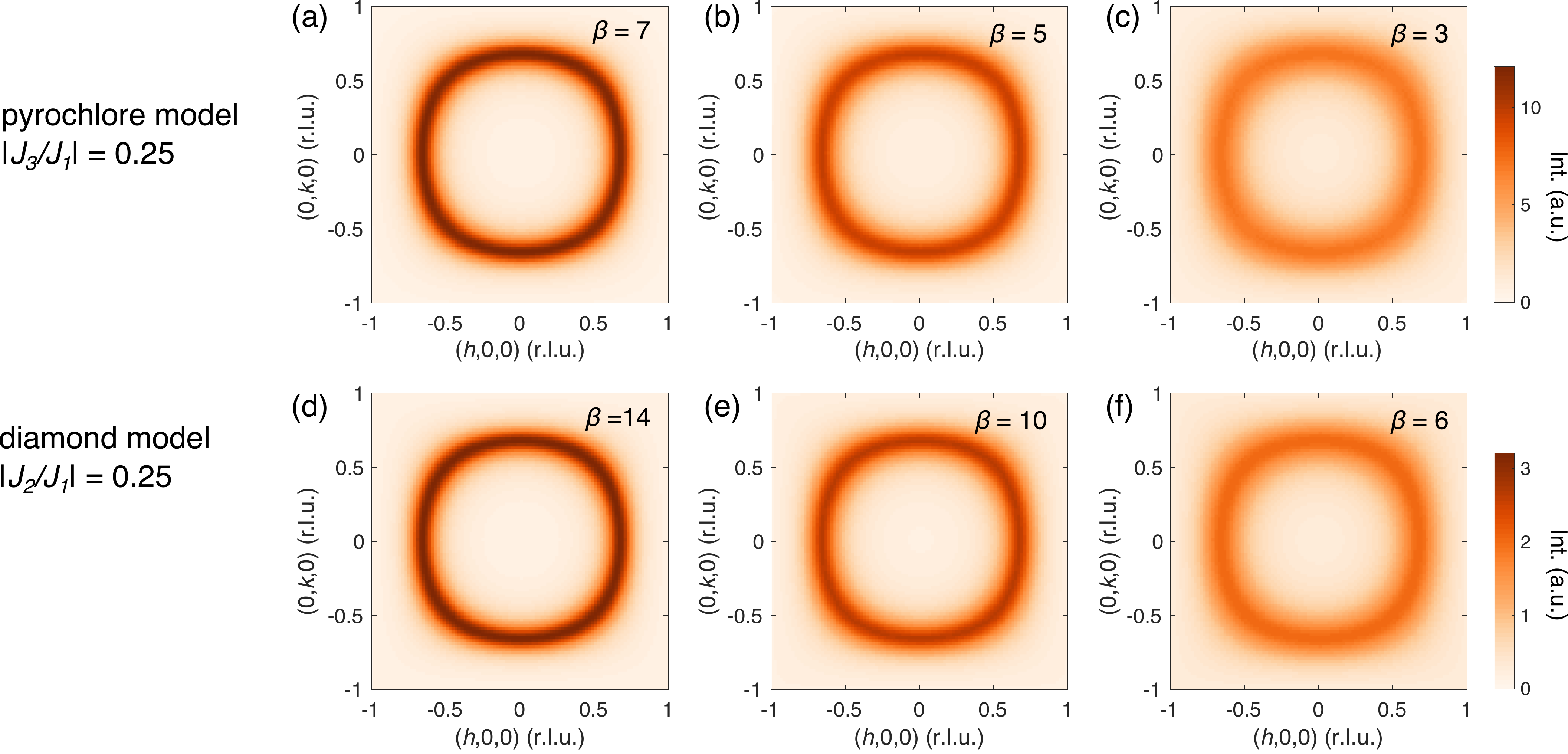}
   \caption{(a-c) Diffuse scattering pattern in the ($hk0$) plane for the $J_1$-$J_3$ model with $|J_3/J_1|=0.25$ at inverse temperature $\beta=7$ (a), 5 (b), and 3 (c). (d-f) Similar calculations for the $J_1$-$J_2$ model with $|J_2/J_1|=0.25$ at $\beta=14$ (a), 10 (b), and 6 (c). Intensities are shown in linear scale.
   \label{fig:scga}}
   \end{figure}

Self-consistent Gaussian approximation (SCGA) calculations were performed to compare the diffuse scattering patterns in the SSL state on the diamond and pyrochlore lattices. Implementation details for our calculations can be found in Ref.~\cite{gao_spiral_2021s}. Figure~\ref{fig:scga} compares the diffuse scattering patterns in the $(hk0)$ plane for the $J_1$-$J_3$ model on a pyrochlore lattice and the $J_1$-$J_2$ model on a diamond lattice. Both the ratio of $|J_3/J_1|$ and $|J_2/J_1|$ are fixed at 0.25. Similar patterns are observed on the pyrochlore and diamond lattice although the degeneracy in the former case is only approximate.

\section{Comparison to the half-moon patterns}

Recent theoretical studies reveal that the pyrochlore and kagome lattices may exhibit half-moon scattering patterns with sub-dimensional degeneracy~\cite{rau_spin_2016s, mizoguchi_magnetic_2018s, hering_fracton_2021s, kiese_pinch_2022s}. These works focus on a $J_1-J_2-J_{3a}$ model with $J_2 = J_{3a}$ and all couplings being AFM. As a contrast, our line-graph model utilizes FM $J_1$, which establishes a different parameter regime to look for sub-dimensional degeneracy and novel spin correlations. Furthermore, our line-graph approach, like the bipartite-lattice models, has the advantage of generating a complete circular ring or sphere in reciprocal space. This contrasts the half-moon patterns discussed for $J_1-J_2-J_{3a}$ model where the scattering ring is separated into two halves. A complete ring indicates an emergent U(1) symmetry in momentum space and is essential in realizing tensor gauge theory~\cite{yan_low_2021s}.

\section{Sample preparation}
Polycrystalline samples of ZnCr$_2$Se$_4$ were synthesized by solid state reaction. Stoichiometric amounts of Zn (99.99\%), Cr (99.95\%), Se (99.99\%), purchased from Alfa Aesar, were ground together inside a glove box and then pressed into a 0.5 inch diameter pellet. The resulting pellet was slowly heated to 600~$^\circ$C and soaked for 10~h. Then the powder was heated to 950~$^\circ$C and held at that temperature for 7~d. The pellet was ground again in Ar-atmosphere and then reheated at 950~$^\circ$C for 3~d. The phase purity of the powder was confirmed by x-ray diffraction collected on a Bruker D2 Phaser diffractometer.

Single crystals of ZnCr$_2$Se$_4$ were grown using the chemical vapor transport (CVT) method with iodine as the transport agent.  Approximately 10 mass percent of iodine was mixed with the polycrystalline sample of ZnCr$_2$Se$_4$ and then sealed in a quartz tube of 19 mm internal diameter. The tube was placed inside a two-zone furnace horizontally and a temperature gradient of approximately 100~$^\circ$C was maintained with the temperature of 950~$^{\circ}$ at the hot end. After eight weeks of growth period, single crystals of average mass $\sim$20~mg were grown at the cold end side of the sealed tube. 

A polycrystalline sample of CuInCr$_4$Se$_8$ was synthesized by directly reacting high purity elements in an SiO$_2$ ampoule sealed under vacuum.  The lowest purity starting material was the Cr (powder) at 99.99~\% metal basis purity.  The ampoule was heated at 50~${^\circ}$C/h to 700~${^\circ}$C for 12~h, then heated to 1100~${^\circ}$C at 50~${^\circ}$C/h for 24~h.  The resulting material was crushed, pressed into a pellet, sealed under vacuum in SiO$_2$ and annealed at 1100~${^\circ}$C for an additional 24~h.  Following this, the product was again ground in air, pressed into a pellet and annealed in a vacuum-sealed SiO$_2$ ampoule for 7~d at 800~${^\circ}$C followed by cooling to room temperature at 200~${^\circ}$C/h.  The resulting powder was examined with powder x-ray diffraction in a PANalytical X'Pert Pro MPD with Cu K$\alpha_1$ radiation.

\section{Sample quality}
\subsection{ZnCr$_2$Se$_4$}
The good quality of our ZnCr$_2$Se$_4$ crystal was confirmed through single crystal x-ray diffraction experiments performed on a Rigaku XtaLAB PRO diffractometer at the Spallation Neutron Source (SNS), Oak Ridge National Laboratory (ORNL). Altogether 5135 reflections, among which 180 are independent, were collected at room temperature with the molybdenum $K$ radiation ($\lambda = 0.71073$~\AA). Refinement of the dataset was performed using the FullProf software~\cite{rodriguez_recent_1993s}.

\begin{figure}[t!]
   \includegraphics[width=0.45\textwidth]{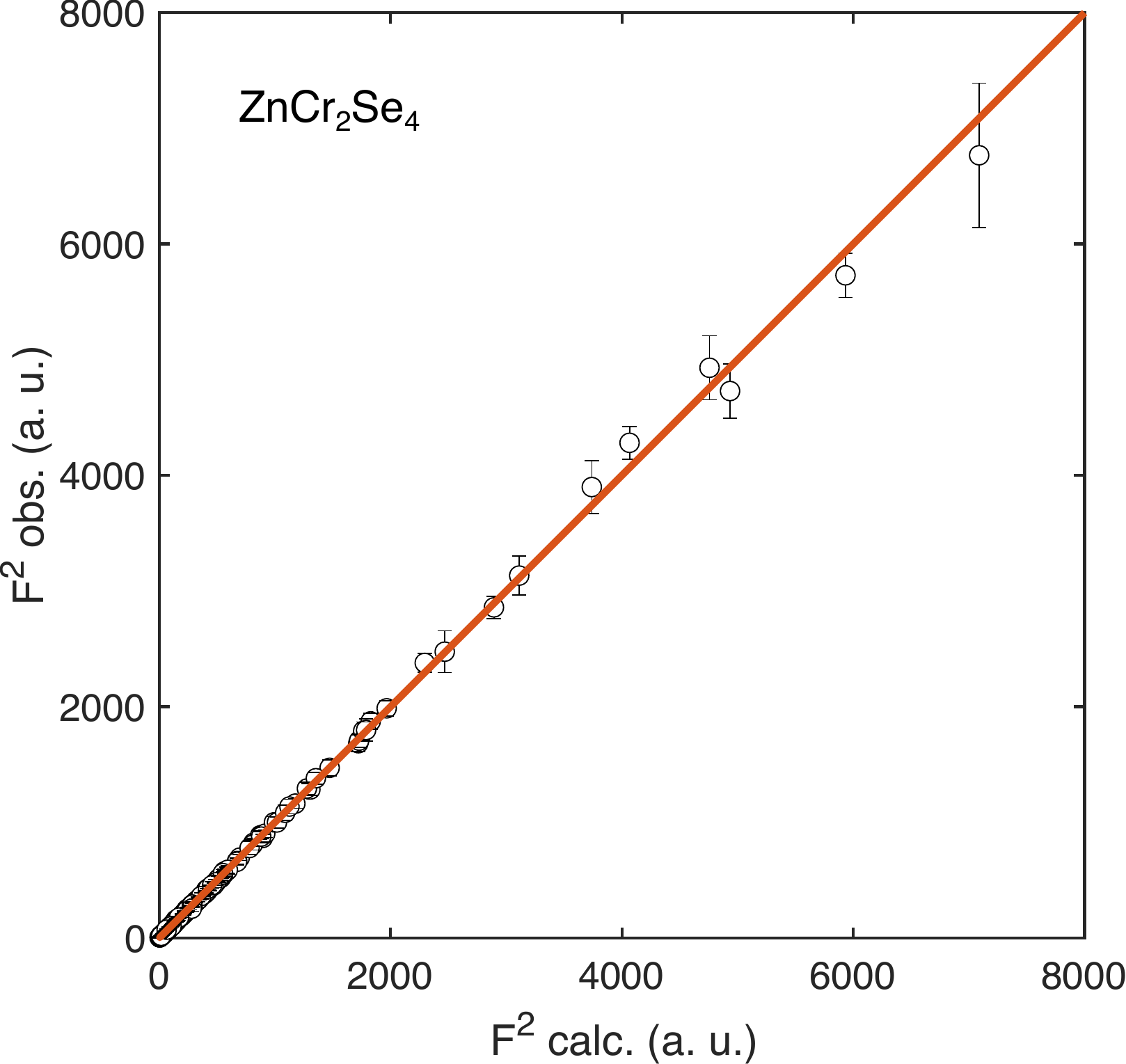}
   \caption{Comparison between the observed and calculated squared structure factor $F^2$ for ZnCr$_2$Se$_4$. 
   \label{fig:refine_ZCSe}}
   \end{figure}

Figure~\ref{fig:refine_ZCSe} compares the observed and calculated squared structure factor $F^2$ assuming the $Fd\overline{3}m$ space group. The lattice constant is determined to be $a = 10.5097(4)$~\AA. The Zn, Cr, Se ions occupy the $8a\,(\frac{1}{8}, \frac{1}{8}, \frac{1}{8})$, $16d\,(\frac{1}{2},\frac{1}{2},\frac{1}{2})$, and $32e\,(x,x,x)$ Wyckoff sites, respectively. The refined Se position is $x = 0.2594(1)$. The goodness-of-fit factors are $R_{F}=1.1$~\% and $R_{F^2}=2.3$~\%. All sites are fully occupied within errors.

\subsection{CuInCr$_4$Se$_8$}
\begin{figure}[t!]
   \includegraphics[width=0.75\textwidth]{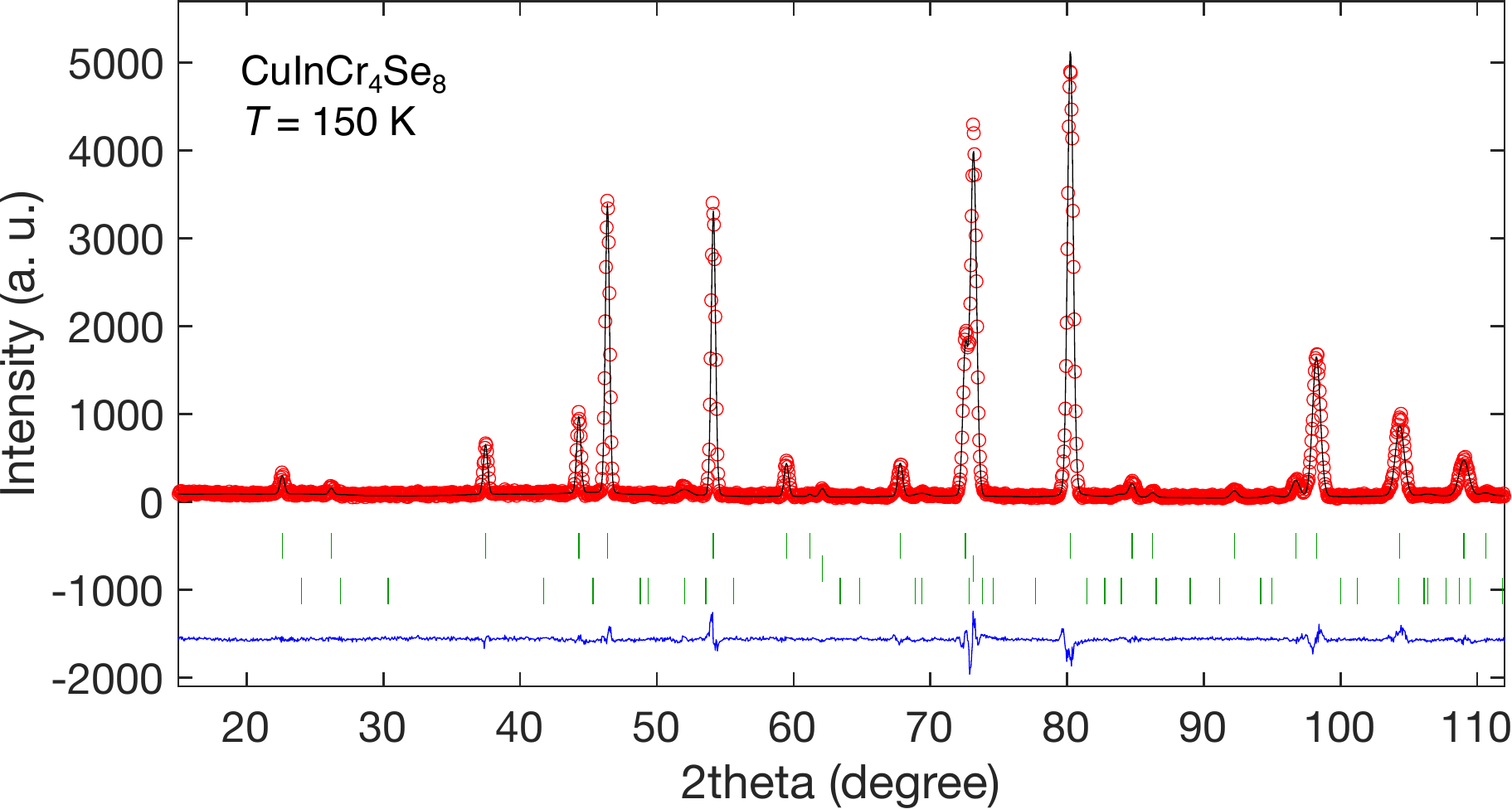}
   \caption{Rietveld analysis of the neutron diffraction data measured at 150~K on HB-2A for a CuInCr$_4$Se$_8$ powder. Data points are shown as red circles. The calculated pattern is shown as the black solid line. Green vertical bars from top to bottom indicate the positions of the Bragg peaks for CuInCr$_4$Se$_8$, aluminum, and Cr$_2$Se$_3$, respectively. The blue line at the bottom shows the difference of measured and calculated intensities.
   \label{fig:refine_CICSe}}
   \end{figure}

The good quality of our CuInCr$_4$Se$_8$ sample was verified through neutron diffraction measurements performed on the HB2A powder diffractometer at the High Flux Isotope Reactor (HFIR), ORNL. Experimental details can be found in the following section of ``Neutron Scattering Experiments". Figure~\ref{fig:refine_CICSe} present the refinement results of the neutron diffraction data collected at $
T = 150$~K. With the cubic space group $F\overline{4}3m$ (\#216), the lattice constant of CuInCr$_4$Se$_8$ was refined to be $a = 10.5550(1)$~\AA. The Cr ions occupy the $16e$ $(x,x,x)$ Wyckoff sites with $x = 0.3669(6)$. The Cu and In ions occupy the $4a$ $(0,0,0)$ and $4d$ $(\frac{3}{4},\frac{3}{4},\frac{3}{4})$ sites, respectively. The Se ions occupy two $16e$ $(x,x,x)$ sites with $x = 0.1322(2)$ and 0.6110(3). The overall goodness-of-fit factor is $\chi^2 = 2.60$, with $R_F=4.8$~\% for the CuInCr$_4$Se$_8$ phase. Bragg peaks at 62$^{\circ}$ and $80.5^{\circ}$ are due to aluminum scattering from the sample holder. Additional Bragg peaks from the impurity phase Cr$_2$Se$_3$ were identified with a refined weight fraction of 3.8~\%. Releasing the site occupancy on the Cu and In sites in the CuInCr$_4$Se$_8$ phase does not improve $\chi^2$, which suggests negligible Cu-In anti-site disorder in our sample. 

\section{Magnetic susceptibility}
\subsection{ZnCr$_2$Se$_4$}
\begin{figure}[h!]
   \includegraphics[width=0.5\textwidth]{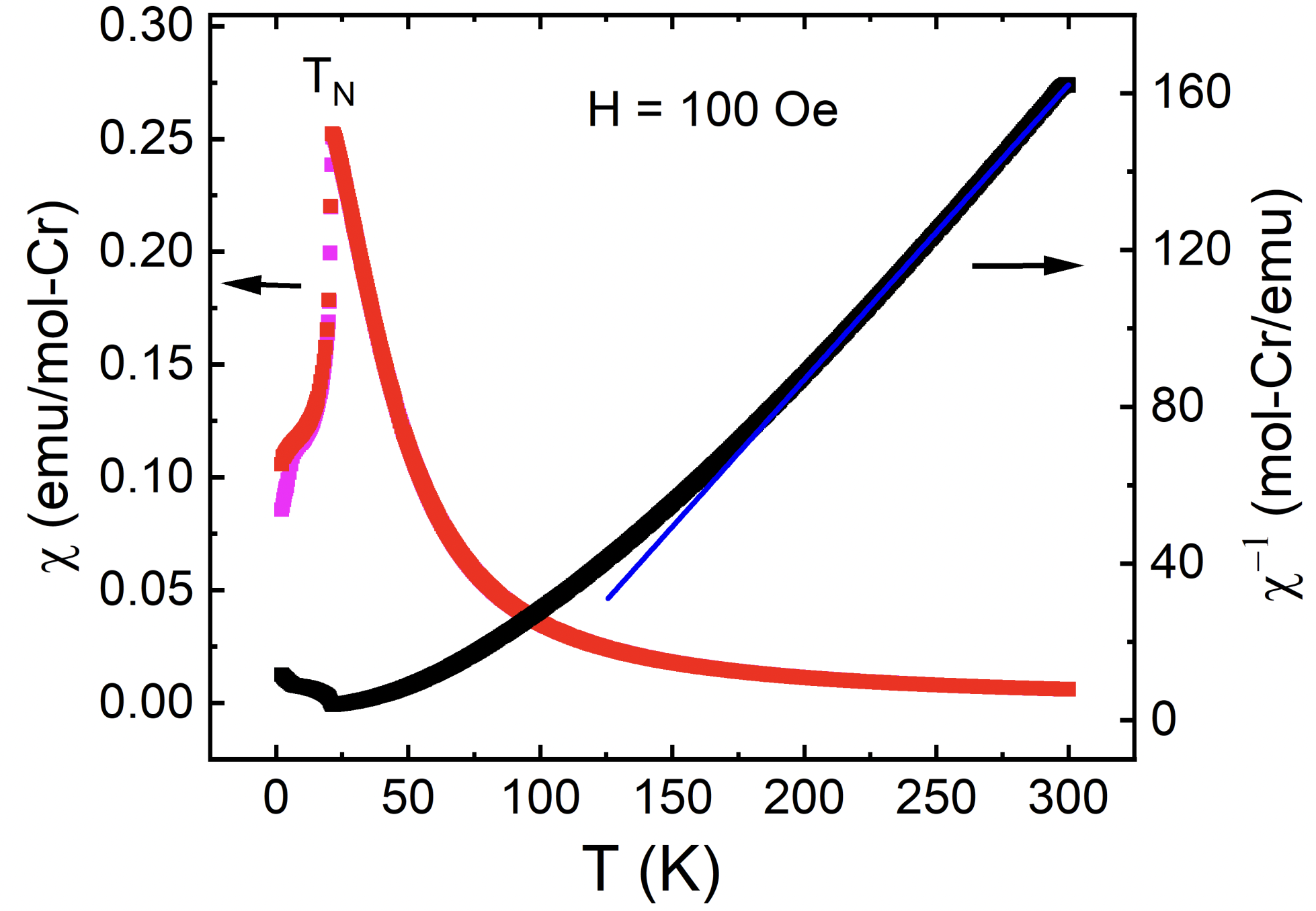}
   \caption{(a) Magnetic susceptibility ($\chi$) and inverse magnetic susceptibility ($\chi^{-1}$) of ZnCr$_2$Se$_4$ measured in 100 Oe. The curves with red and pink squares represent the field-cooled and zero field-cooled magnetic susceptibilities, respectively. Note that these curves are nearly indistinguishable on this figure except for temperatures below 20 K. The blue color solid line represents a Curie-Weiss fit to $\chi^{-1}(T)$ above 200 K. 
   \label{fig:chiT_zcse}}
   \end{figure}

Magnetization measurements on a ZnCr$_2$Se$_4$ crystal were carried out with a Quantum Design Magnetic Property Measurement System. The measuring field was applied perpendicular to the (111) direction. The magnetic susceptibility ($\chi$) and inverse magnetic susceptibility ($\chi^{-1}$) measured in 100~Oe are shown in Fig.~\ref{fig:chiT_zcse}. $\chi(T)$ shows a sharp drop around 21 K, consistent with an AFM LRO transition. The observed transition temperature ($T_N$ $\sim$ 21 K) is consistent with previous literature \cite{hemberger_large_2007s, cameron_magnetic_2016s, izabela_study_2021s}. The zero-field-cooled and field-cooled magnetic susceptibilities overlap with each other.   $\chi^{-1}(T)$ shows linear temperature dependence, Curie-Weiss behavior, above 200 K. Fitting in the temperature range of [200, 300]~K yields an effective
moment of 3.4 $\mu_B$ and a Curie-Weiss temperature of $\sim 85$ K, both being comparable with the previous reports~\cite{hemberger_large_2007s,izabela_study_2021s}. 

\subsection{CuInCr$_4$Se$_8$}
The magnetization of a powder sample of CuInCr$_4$Se$_8$ was characterized using a Quantum Design Magnetic Property Measurement System and the ac magnetic susceptibility was measured using a Quantum Design Physical Property Measurement System with further ac measurements performed in a Quantum Design MPMS3. 

\begin{figure}[h!]
   \includegraphics[width=0.8\textwidth]{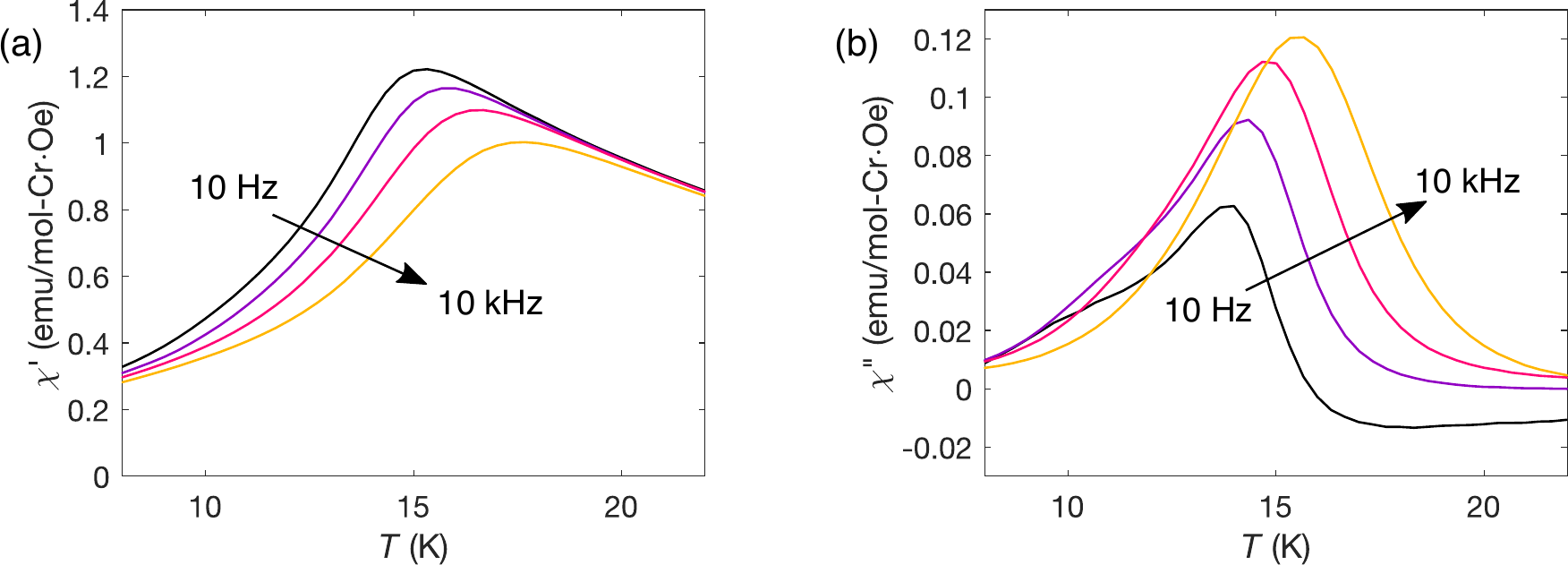}
   \caption{Real (a) and imaginary (b) parts of the ac magnetic susceptibility for CuInCr$_4$Se$_8$ measured in a 10 Oe field. 
   \label{fig:CICS_ac}}
   \end{figure}

Figure~\ref{fig:CICS_ac} presents the real and imaginary parts of the ac magnetic susceptibility for CuInCr$_4$Se$_8$ measured in a 10~Oe field at frequencies of 10, 100, 1000, and 10000~Hz. The real part is also shown in Fig.~4(a) of the main text. The freezing temperature increases with increasing frequencies, exhibiting a typical spin glass behavior.

\section{Neutron scattering experiments}
Diffuse neutron scattering experiments for ZnCr$_2$Se$_4$ were performed on the CORELLI elastic diffuse scattering spectrometer~\cite{ye_implementation_2018s} at the SNS, ORNL. Like a typical time-of-flight (TOF) direct geometry spectrometer, the energy resolution depends on the incident neutron energy $E_i$. For $E_i = 13\sim33$~meV that is related to our scattering range, the average energy resolution is estimated to be $\sim0.8$~meV. A crystal (mass $\sim80$~mg) was aligned with the $c$ axis vertical. A helium flow cryostat was employed to reach temperatures $T$ down to $\sim1.4$ K. Measurements below $T_N$ were performed at 20~K to suppress strong Bragg scattering. Data were acquired by rotating the sample in 1$^{\circ}$ steps, covering a total range of 30$^{\circ}$. Data reduction and projection were performed using the MANTID software~\cite{arnold_mantid_2014s}. SCGA fits of the diffuse scattering data utilize the combined simulated annealing and simplex optimization methods as implemented in the iFit package~\cite{farhi_ifit_2014s, gao_spiral_2021s}.

Neutron diffraction experiments on CuInCr$_4$Se$_8$ were performed on the HB-2A diffractometer at the HFIR, ORNL~\cite{calder_suite_2018s}. About 5~g powder sample was filled in an aluminum sample can. An incident neutron wavelength of 2.41~\AA\ was selected by the Ge-113 monochromator, and the collimators are set as open-open-12$^{\prime}$. A $^3$He closed cycle refrigerator (CCR) was employed to reach temperatures down to 0.25~K. Counting time was 12 hrs at each temperature. Additional measurements were performed using an orange cryostat with a base temperature of 1.5~K.

Inelastic neutron scattering experiments on CuInCr$_4$Se$_8$ were performed on the SEQUOIA spectrometer at SNS, ORNL. The same powder sample as was used for the diffraction measurements was placed in an aluminum can. A CCR was employed to reach temperatures down to 5 K. The incident neutron energy was selected at $E_i$ = 10, 25, and 50~meV, with a corresponding Fermi chopper frequency of 180, 240, 360~Hz, respectively, in the high-resolution setup. Data from an empty sample can were subtracted as the background. Data reduction was performed using the MANTID software~\cite{arnold_mantid_2014s}. SCGA fits of the integrated equal-time correlations utilize the combined simulated annealing and simplex optimization methods as that for ZnCr$_2$Se$_4$.

\begin{figure}[b!]
   \includegraphics[width=0.9\textwidth]{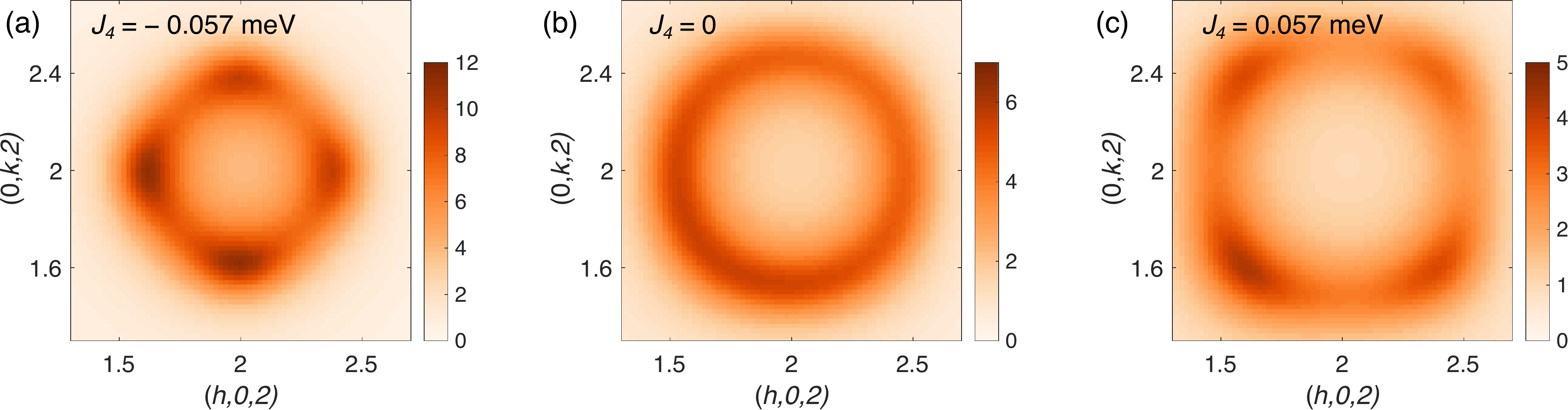}
   \caption{(a) Calculated diffuse scattering pattern in the ($hk2$) plane at $T = 30$~K for the $J_1$-$J_2$-$J_3$-$J_4$ model on a pyrochlore lattice. The magnetic form factor of the Cr$^{3+}$ ion is applied. The parameter set is the same as that in Fig.~3 of the main text,  which includes a nonzero $J_4 = -0.057$~meV. (b,c) Diffuse scattering patterns expected at zero $J_4$ (b) and positive $J_4=0.057$~meV (c). Intensities are shown in linear scale.
   \label{fig:J4}}
   \end{figure}

\section{Perturbations from $J_4$ in Z\lowercase{n}C\lowercase{r}$_2$S\lowercase{e}$_4$}

As discussed in the main text, the slight shrinking of the spiral surface and the intensity modulation over the spiral surface is mainly due to the perturbation of a FM $J_4$ interaction. Figure~\ref{fig:J4} compares the diffuse pattern in the ($hk2$) plane calculated by the SCGA method for FM, zero, and AFM $J_4$ perturbations. Without $J_4$, intensity over the scattering ring is only weakly modulated by the Cr$^{3+}$ magnetic form factor. While for FM and AFM $J_4$ perturbations, stronger intensities are observed along the (100) and (110) directions, respectively. The former case is observed in ZnCr$_2$Se$_4$, while the latter case is similar to the AFM $J_3$ perturbation in MnSc$_2$S$_4$~\cite{gao_spiral_2017s}.

\begin{figure}[b!]
   \includegraphics[width=0.9\textwidth]{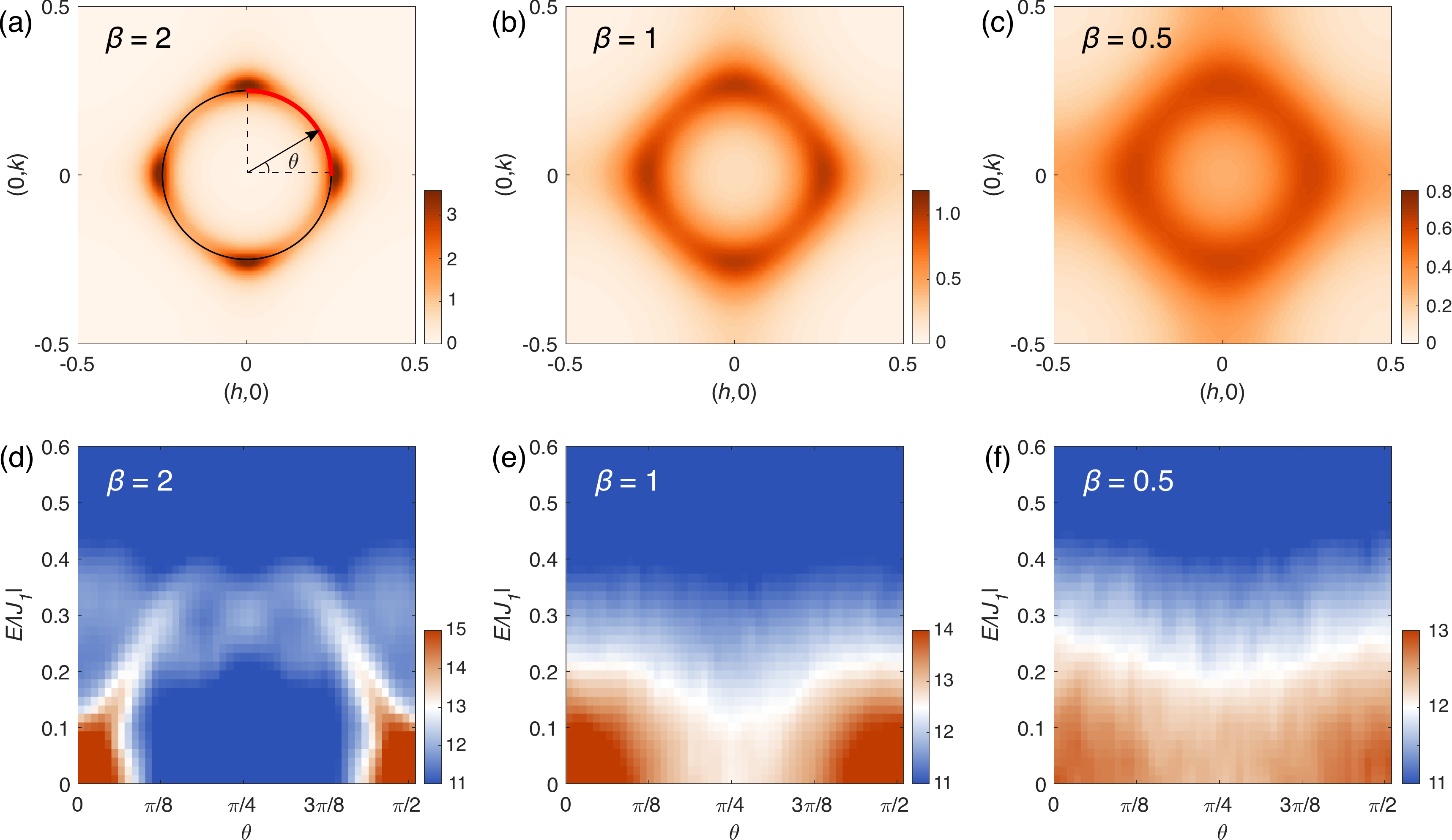}
   \caption{(a-c) Diffuse scattering patterns for a square-lattice model with couplings $J_1 = -1$, $J_2 = 2J_3 = 0.5$, and $J_4 = 0.05$. The inverse temperatures are $\beta = 2$ (a), 1 (b), and 0.5 (c). Intensities are shown in linear scale. (d-f) Low-energy INS spectra for the same square-lattice model at inverse temperatures of $\beta = 2$ (d), 1 (e), and 0.5 (f). The scanning path in reciprocal space is a quarter of the circular ring with $\theta$ defined in panel (a). Intensities are shown in log scale. 
   \label{fig:square}}
   \end{figure}

It is noteworthy that although the perturbation of $J_4$ induces a magnetic long-range order with dispersive magnon excitations as previously studied using INS~\cite{tymoshenko_pseudo_2017s}, the excitation gap observed along the spiral surface will disappear at temperatures above $T_N$ due to incoherent spin excitations, which justifies an emergent SSL state. As an illustration, we calculate the spin correlations of a simplified SSL model on a square lattice with $J_1 = -1$, $J_2 = 2J_3 = 0.5$, and a perturbational $J_4 = 0.05$. In the case of $J_4 = 0$, this model exhibits a perfect spiral spin liquid phase~\cite{yan_low_2021s}. The presence of AFM $J_4$ perturbations induces a tendency to a spiral long-range order with $\bm{q} = (0.25, 0)$ at $\beta \approx 1.2$. Figure~\ref{fig:square}(a-c) summarizes the diffuse scattering patterns in the $(h,k)$ plane calculated by the SCGA method at $\beta= 2$, 1, and 0.5. Figure~\ref{fig:square}(d-f) summarizes the corresponding low-energy INS spectra calculated by the molecular dynamics method~\cite{pohle_theory_2021s} using a superlattice of $80\times80$. The scanning path is along the scattering ring in reciprocal space as indicted by the red curve in Fig.~\ref{fig:square}(a). In the nearly ordered phase at $\beta= 2$, gapped excitations are observed along the scattering ring at wavevector transfers away from the ordered $\bm{q}$ positions. However, at $\beta = 1$ and 0.5 in the paramagnetic phase, the excitation gap disappears, meanwhile the diffuse scattering pattern become a complete circular ring. In ZnCr$_2$Se$_4$, we expect the weak excitation gap along the spiral surface observed in the ordered phase~\cite{tymoshenko_pseudo_2017s} to be similarly removed at temperatures above $T_N$, so that the quasielastic spin dynamics become dominant over the entire spiral surface in the SSL state.

\begin{figure}[b!]
   \includegraphics[width=0.95\textwidth]{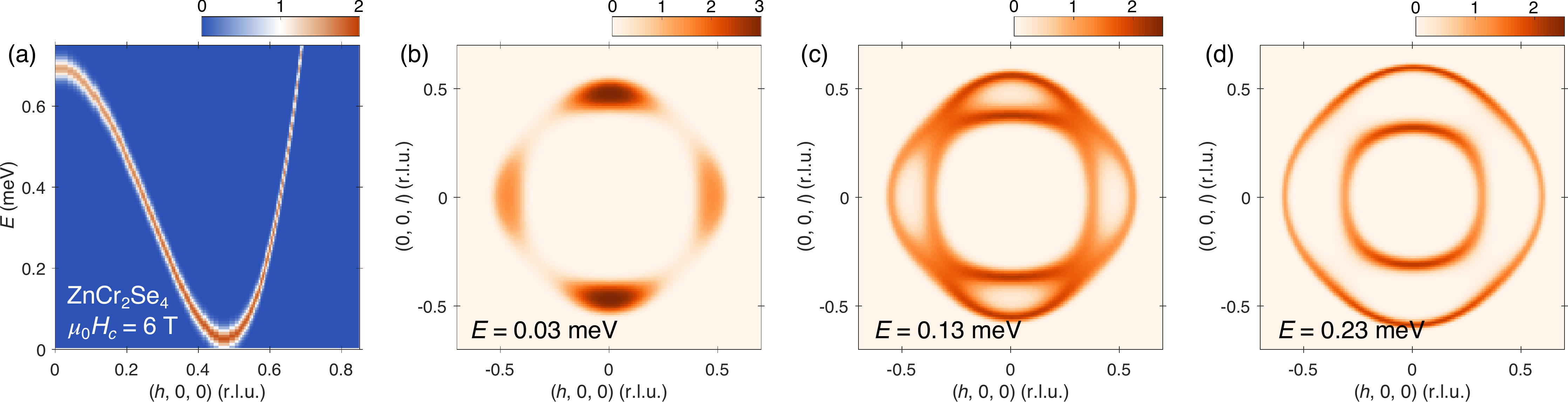}
   \caption{(a) Calculated magnon dispersion in the field-induced FM phase of ZnCr$_2$Se$_4$. The coupling strengths are the same as reported in Ref.~\cite{tymoshenko_pseudo_2017s}, and a 6 T field is applied along the $c$ axis. (b-d) Constant-energy slices of the magnon spectra at $E = 0.03$ (b), 0.13 (c), and 0.23~meV (d) with an integration width of 0.02~meV. Intensities are shown in linear scale.
   \label{fig:field}}
   \end{figure}

\section{Field-induced excitations in Z\lowercase{n}C\lowercase{r}$_2$S\lowercase{e}$_4$}

The magnon excitations in ZnCr$_2$Se$_4$ have been studied using INS below $T_N$ in a magnetic field along the $c$ axis~\cite{inosov_magnetic_2020s}. It was observed that in a 6 T field, the magnon bands soften almost isotropically in reciprocal space, leading to ring-like scattering at low energies. By assuming a FM long-range ordered ground state~\cite{hemberger_large_2007s}, the main features of the low-energy excitations can be reproduced by linear spin wave calculations~\cite{toth_linear_2015s} as shown in Fig.~\ref{fig:field}. Especially, the experimentally observed scattering ring~\cite{inosov_magnetic_2020s}, including the four empty holes around the long-range order $\bm{q}$, is captured in the constant-energy slice at $E$ = 0.13 meV shown in Fig.~\ref{fig:field}(c). The four empty holes arise from the dispersion of the softened magnons, and the strong intensities around $\bm{q} = \{0, 0, 0.47\}$ at lower energies shown in Fig.~\ref{fig:field}(b) indicate that the quasielastic dynamics is dominated by fluctuations of discrete $\bm{q}$ vectors. Although both observations contradict the scenario of a field-induced SSL state in ZnCr$_2$Se$_4$, the rather close energy difference over the spiral surface (compared to the total magnon bandwidth of $\sim 33$~meV) still suggests magnetic field as a possible route to SSLs, which is a very interesting route for future research.

\begin{figure}[h!]
   \includegraphics[width=0.9\textwidth]{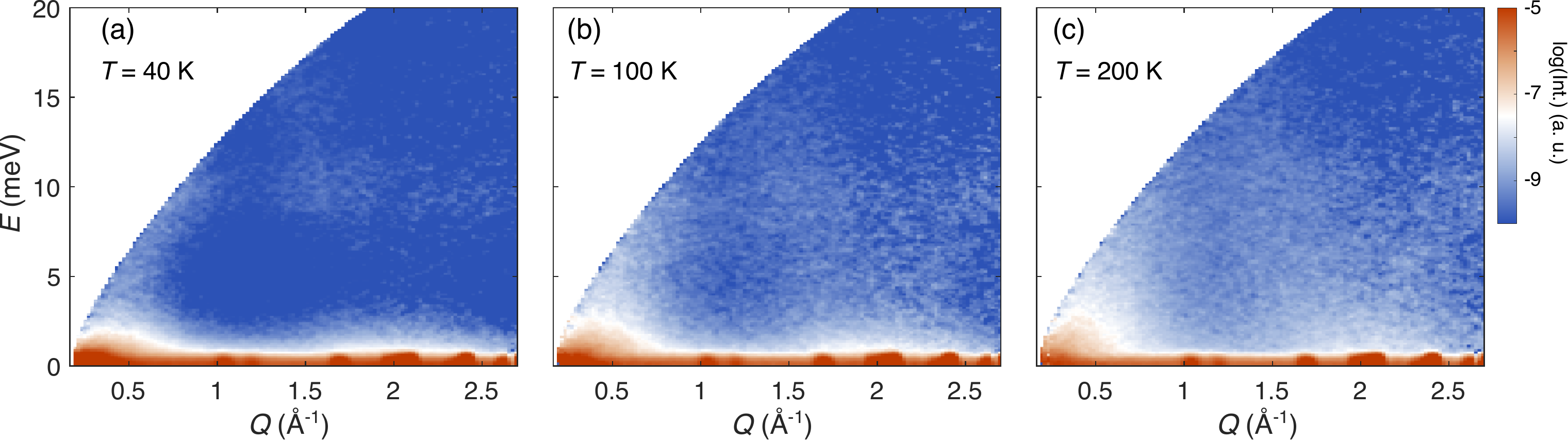}
   \caption{INS spectra $S(Q,\omega)$ measured on SEQUOIA with an incident neutron energy of $E_i=25$~meV. The measuring temperatures are $T = 40$~K (a), 100~K (b), and 200~K (c). 
   \label{fig:seq}}
   \end{figure}

\section{INS spectra for C\lowercase{u}I\lowercase{n}C\lowercase{r}$_4$S\lowercase{e}$_8$}

Figure~\ref{fig:seq} presents the INS spectra $S(Q,\omega)$ measured on SEQUOIA with an incident neutron energy of $E_i=25$~meV. The measuring temperatures are $T = 40$~K (a), 100~K (b), and 200~K (c). Empty can data collected at the corresponding temperatures have been subtracted as the background. The equal-time spin correlations shown in Fig.~4(c-e) of the main text are obtained by integrating the INS spectra from $[-20, 20]$~meV.

\section{General guidelines for SSL search in real materials}

Several general guidelines for the realization of spiral spin liquids on the line-graph lattices can be drawn from the example compound ZnCr$_2$Se$_4$ and other known candidates. Firstly, the favorable electron configurations should minimize the single-ion anisotropy so that spiral correlations along all directions can be stabilized.  Examples include some instances of the d$^3$ and d$^5$ configurations. Secondly, since the line-graph approach relies on ferromagnetic $J_1$ and antiferromagnetic $J_3$ couplings, qualitative analysis of the coupling signs through the Goodenough-Kanamori rule should be applied as an initial check for promising compounds. For example, in ZnCr$_2$Se$_4$, the $J_1$ couplings are ferromagnetic because of the $90^{\circ}$ Cr$^{3+}$-O$^{2-}$-Cr$^{3+}$ super-exchange path, while the $J_3$ couplings are antiferromagnetic because the dominant super-exchange terms are those between two empty or two half-filled orbitals. The $J_2$ strength, which should be negligible in the line-graph approach, is difficult to predict in an empirical way as it involves competing $e_g$-$t_{2g}$ and $t_{2g}$-$t_{2g}$ exchange terms. Considering the weak $J_2$ strength in ZnCr$_2$Se$_4$ has been successfully reproduced by DFT calculations~\cite{yaresko_electronic_2008s}, we expect similar calculations can be employed for compounds of interests to guide further experimental studies.

Aside from empirical conjectures and theoretical calculations, magnetic orders determined from previous neutron diffraction experiments (for example cataloged at the Bilbao crystallographic server~\cite{aroyo_crystal_2011s}) may also inform the material search. This is because the ordering transition in spiral spin liquids, no matter due to the order-by-disorder effect or further-neighbor perturbations, often leads to a spiral magnetic structure. Therefore, examining line-graph lattice compounds known with a spiral long-range order will be a promising strategy to identify additional examples of materials hosting a spiral spin liquid.

\end{document}